\documentclass[preprints,article,accept,moreauthors,pdftex]{mdpi}

\usepackage[utf8]{inputenc}

\usepackage{nicefrac}
\usepackage{siunitx}

\usepackage{graphicx}
\usepackage{amsmath}
\usepackage{xcolor}
\usepackage[normalem]{ulem}

\usepackage{hyperref}

\definecolor{linkcolor}{rgb}{0,0.3,0.7}
\hypersetup{colorlinks=true,allcolors=linkcolor}

\def\vf{v_\mathrm{f}}
\def\vp{v_\mathrm{p}}
\def\snc{s_\mathrm{nc}}

\let\Re\relax
\DeclareMathOperator{\Re}{Re}
\let\Im\relax
\DeclareMathOperator{\Im}{Im}

\DeclareMathOperator{\tr}{tr}

\firstpage{1} 
\pubvolume{xx}
\issuenum{1}
\articlenumber{5}
\pubyear{2020}
\copyrightyear{2020}
\history{}

\Title{Cytosolic flow induced symmetry breaking in a conceptual polarity model}

\Author{Manon C. Wigbers$^{\dagger}$ , Fridtjof Brauns$^{\dagger}$ , Ching Yee Leung$^{\dagger}$ and Erwin Frey*}

\AuthorNames{Manon C. Wigbers, Fridtjof Brauns, Ching Yee Leung and Erwin Frey}

\address[1]{%
Arnold Sommerfeld Center for Theoretical Physics (ASC) and Center for NanoScience (CeNS), Department of Physics, Ludwig-Maximilians-Universit\"at M\"unchen, Theresienstra\ss e 37, D--80333 M\"unchen, Germany
}

\corres{Correspondence: frey@lmu.de;}

\firstnote{These authors contributed equally to this work.} 

\abstract{Important cellular processes, such as cell motility and cell division, are coordinated by cell polarity, which is determined by the non-uniform distribution of certain proteins. 
Such protein patterns form via an interplay of protein reactions and protein transport.
Since Turing's seminal work, the formation of protein patterns resulting from the interplay between reactions and diffusive transport has been widely studied.
Over the last few years, increasing evidence shows that also advective transport, resulting from cytosolic and cortical flows, is present in many cells. 
However, it remains unclear how and whether these flows contribute to protein-pattern formation. 
To address this question, we use a minimal model that conserves the total protein mass to characterize the effects of cytosolic flow on pattern formation. 
Combining a linear stability analysis with numerical simulations, we find that membrane-bound protein patterns propagate \emph{against} the direction of cytoplasmic flow with a speed that is maximal for intermediate flow speed.
We show that the mechanism underlying this pattern propagation relies on a higher protein influx on the upstream side of the pattern compared to the downstream side.
Furthermore, we find that cytosolic flow can change the membrane pattern qualitatively from a peak pattern to a mesa pattern.
Finally, our study shows that a non-uniform flow profile can induce pattern formation by triggering a regional lateral instability. }


\begin{document}

\section{Introduction}
Many biological processes rely on the spatiotemporal organization of proteins.
Arguably one of the most elementary forms of such organization is cell polarization --- the formation of a ``cap'' or spot of high protein concentration that determines a direction.
Such a polarity axis then coordinates downstream processes including motility \cite{Weiner2002,Keilberg.Sogaard-Andersen2014}, cell division~\cite{Bi.Park2012}, and directional growth~\cite{Chiou.etal2017}.
Cell polarization is an example for \emph{symmetry breaking}~\cite{Goryachev.Leda2017}, as the orientational symmetry of the initially homogeneous protein distribution is broken by the formation of the polar cap.

Intracellular protein patterns arise from the interplay between protein interactions (chemical reactions) and protein transport. 
Diffusion in the cytosol serves as the most elementary means of transport. 
Pattern formation resulting from the interplay of reactions and diffusion has been widely studied since Turing's seminal work~\cite{Turing1952}.
In addition to diffusion, proteins can be transported by fluid flows in the cytoplasm and along cytoskeletal structures (vesicle trafficking, cortical contractions) driven by molecular motors~\cite{Hawkins.etal2009,Calvez.etal2020}. 
These processes lead to \emph{advective} transport of proteins.

Recently, it has been shown experimentally that advective transport (caused by cortical flows) induces polarization of the PAR system in the \emph{C. elegans} embryo \cite{Goehring.etal2011,Illukkumbura.etal2020,Gross.etal2019}.
Furthermore, \emph{in vitro} studies with the MinDE system of \emph{E.~coli}, reconstituted in microfluidic chambers, have shown that the flow of the bulk fluid has a strong effect on the protein patterns that form on the membrane~\cite{Ivanov.Mizuuchi2010,Vecchiarelli.etal2014}.
Increasing evidence shows that cortical and cytosolic flows (also called ``cytoplasmic streaming'') are present in many cells \cite{Grill.etal2001,Munro.etal2004,Hecht.etal2009,Mayer.etal2010,Goldstein.vandeMeent2015,Mogilner.Manhart2018}. In addition, cortical contractions can drive cell-shape deformations \cite{Bischof.etal2017}, inducing flows in the incompressible cytosol \cite{Koslover.etal2017,Klughammer.etal2018}.
However, the role of flows for protein-pattern formation remains elusive.
This motivates to study the role of advective flow from a conceptual perspective, with a minimal model.
The insights thus gained will help to understand the basic, principal effects of advective flow on pattern formation and reveal the underlying elementary mechanisms.

The basis of our study is a paradigmatic class of models for cell polarization that describe a single protein species which has a membrane-bound state and a cytosolic state.
Such two-component mass-conserving reaction--diffusion (2cMcRD) systems serve as conceptual models for cell polarization~\cite{Mori.etal2008, Jilkine.Edelstein-Keshet2011, Diegmiller.etal2018, Chiou.etal2018, Brauns.etal2018}. Specifically they have been used to model Cdc42 polarization in budding yeast~\cite{Otsuji.etal2007,Goryachev.Pokhilko2008} and PAR-protein polarity~\cite{Trong.etal2014}.
2cMcRD systems generically exhibit both spontaneous and stimulus-induced polarization~\cite{Trong.etal2014,Goryachev.Leda2017,Brauns.etal2018}.
In the former case, a spatially uniform steady state is unstable against small spatial perturbations (``Turing instability''~\cite{Turing1952}).
Adjacent to the parameter regime of this lateral instability, a sufficiently strong, localized stimulus (e.g.\ an external signal) can induce the formation of a pattern starting from a stable spatially uniform state. 
The steady state patterns that form in two-component McRD systems are generally stationary (there are no travelling or standing waves).
Moreover, the final stationary pattern has no characteristic wavelength.
Instead, the peaks that grow initially from the fastest growing mode (``most unstable wavelength'') compete for mass until only a single peak remains (``winner takes all'')~\cite{Ishihara.etal2007,Otsuji.etal2007,Chiou.etal2018}.
The location of this peak can be controlled by external stimuli (e.g.\ spatial gradients in the reaction rates)~\cite{Otsuji.etal2007,Wigbers.etal2020}.

Recently, a theoretical framework, termed \emph{local equilibria theory}, has been developed to study these phenomena using a geometric analysis in the phase plane of the protein concentrations~\cite{Brauns.etal2018,Halatek.etal2018}. 
With this framework one can gain insight into the mechanisms underlying the dynamics of McRD systems both in the linear and in the strongly nonlinear regime, thereby bridging the gap between these two regimes.

Here, we show that cytosolic flow in two-component systems always induces upstream propagation of the membrane-bound pattern. In other words, the peak moves against the cytosolic flow direction. 
This propagation is driven by a higher protein influx on the upstream side of the membrane-concentration peak compared to its downstream side. 
Using this insight, we are able to explain why the propagation speed becomes maximal at intermediate flow speeds and vanishes when the rate of advective transport becomes fast compared to the rate of diffusive transport or compared to the reaction rates.
We first study a uniform flow profile using periodic boundaries.
This effectively represents a circular flow, which is observed in plant cells (where this phenomenon is called cytoplasmic streaming or cyclosis)~\cite{Allen.Allen1978}.
It also represents an \emph{in vitro} system in a laterally large microfluidic chamber.
We then study the effect of a spatially non-uniform flow profile in a system with reflective boundaries, as a minimal system for flows close to the membrane~\cite{Goehring.etal2011,Gross.etal2019,Diegmiller.etal2018}, e.g.\ in the actin cortex.
We show that a non-uniform flow profile redistributes the protein mass, which can trigger a regional lateral instability and thereby induce pattern formation from a stable homogeneous steady state.

The remainder of the paper is structured as follows. 
We first introduce the model in Sec. \ref{sec:model}.
We then perform a linear stability analysis in Sec.~\ref{sec:stability} to show how spatially uniform cytosolic flow influences the dynamics close to a homogeneous steady state. 
In Sec.~\ref{sec:steady-state}, we use numerical simulations to study the fully nonlinear long-term behavior of the system. 
Next, we show that upon increasing the cytosolic flow velocity, the pattern can qualitatively change from a mesa pattern to a peak pattern in Sec.~\ref{sec:mesa-peak}.
Finally, in Sec.~\ref{sec:pattern-trigger}, we study how a spatially non-uniform cytosolic flow can trigger a regional lateral instability and thus induce pattern formation.
Implications of our findings and links to earlier literature are briefly discussed at the end of each section. We conclude with a brief outlook section.

\section{Model} \label{sec:model}

\begin{figure}
  \centering
  \includegraphics{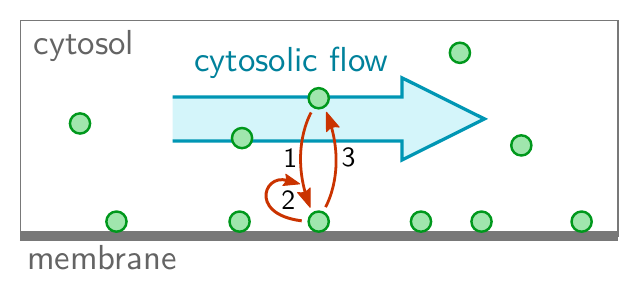}
  \caption{One-dimensional two-component system with cytosolic flow into the positive $x$ direction. The reaction kinetics include (1) attachment, (2) self-recruitment and (3) enzyme-driven detachment.
  }
  \label{fig:1}
\end{figure}

We consider a spatially one-dimensional system of length $L$.
The proteins can cycle between a membrane-bound state (concentration $m(x,t)$) and a cytosolic state (concentration $c(x,t)$), and diffuse with diffusion constants $D_m$ and $D_c$, respectively (Fig.~\ref{fig:1}). 
In cells, the diffusion constant on the membrane is typically much smaller than the diffusion constant in the cytosol. 
In the cytosol, the proteins are assumed to be advected with a speed $\vf(x)$, as indicated by the blue arrow in Fig.~\ref{fig:1}. 
Thus, the reaction-diffusion-advection equations for the cytosolic density and membrane density read 
\begin{subequations} \label{eq:model}
\begin{align}
    \partial_t c + \partial_x(\vf c) &= \kern0.34em D_c \partial_x^2 c \kern0.5em - f(m,c) , \\
    \partial_t m &= D_m \partial_x^2 m + f(m,c) ,
\end{align}
\end{subequations}
with either periodic or reflective boundary conditions.
The nonlinear function $f(m,c)$ describes the reaction kinetics of the system.
Attachment--detachment kinetics can generically be written in the form 
\begin{equation} \label{eq:att-det-kinetics}
    f(m,c) = a(m) c - d(m) m,
\end{equation}
where $a(m) > 0$ and $d(m) > 0$ denote the rate of attachment from the cytosol to the membrane and detachment from the membrane to the cytosol, respectively.
The dynamics given by Eq.~\eqref{eq:model} conserve the average total density
\begin{equation}
	 \bar{n} = \frac{1}{L} \int_0^L \! \mathrm{d}x \; n (x,t).
\end{equation}
Here, we introduced the \emph{local} total density $n(x,t) := m(x,t) + c(x,t)$.

For illustration purposes, we will use a specific realization of the reaction kinetics~\cite{Brauns.etal2018},
\begin{equation}
    a(m) = k_\mathrm{on} + k_\mathrm{fb} m \quad \text{and} \quad
    d(m) = \frac{k_\mathrm{off}}{K_\mathrm{D} + m},
\end{equation}
describing attachment with a rate $k_\mathrm{on}$, self-recruitment with a rate $k_\mathrm{fb}$, and enzyme-driven detachment with a rate $k_\mathrm{off}$ and the Michaelis-Menten constant $K_\mathrm{D}$, respectively. 
However, our results do not depend on the specific choice of the reaction kinetics.
Unless stated otherwise, we use the parameters: $k_\mathrm{on}=\SI{1}{s^{-1}}, k_\mathrm{fb}=\SI{1}{\micro m.s^{-1}}, k_\mathrm{off}=\SI{2}{s^{-1}}, K_\mathrm{D}=\SI{1}{\micro m^{-1}}, \bar{n}=\SI{5}{\micro m^{-1}}, D_m=\SI{0.01}{\micro m^2\!/s}, D_c=\SI{10}{\micro m^2\!/s}$.

\section{Linear stability analysis}
\label{sec:stability}

\begin{figure}
  \centering
  \includegraphics{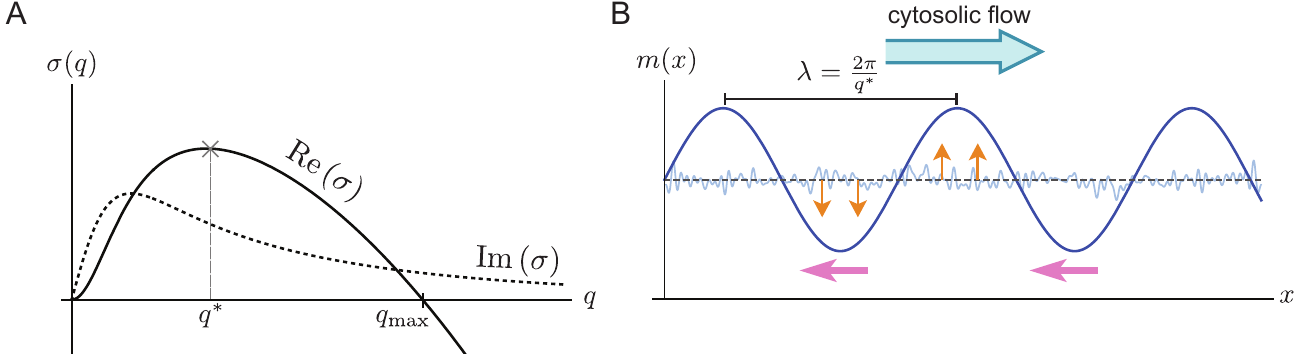}
  \caption{
  (A) Sketch of real (solid) and imaginary (dotted) part of a typical dispersion relation with a band $[0,q_\mathrm{max}]$ of unstable modes.
  (B) The initial dynamics of a spatially homogeneous state with a small random perturbation (blue thin line). The direction of cytosolic flow is indicated by a blue arrow. The typical wavelength ($\lambda$) of the initial pattern is determined by the fastest growing mode $q^*$ and the phase velocity is determined by the value of the imaginary part of dispersion relation at the fastest growing mode ($v_\text{phase} = \text{Im}\sigma(q^*)/{q^*}$). The growth of the pattern is indicated by orange arrows, while the travelling direction is indicated by pink arrows. 
  }
\label{fig:2}
\end{figure}

\subsection{Linearized dynamics and basic results}

To study how cytosolic flow affects the formation of protein patterns, we first consider a spatially uniform flow profile (i.e.\ constant $\vf(x) = \vf$) and perform a linear stability analysis of a spatially homogeneous steady state $\mathbf{u}^* = (m^*,c^*)$:
\begin{equation} \label{eq:HSS}
    f(m^*,c^*) = 0, \quad m^* + c^* = \bar{n}.
\end{equation}
Following the standard procedure, we linearize the dynamics for small perturbations $\mathbf{u}(x,t) = \mathbf{u}^* + \delta \mathbf{u}(x,t)$ around the homogeneous steady state. 
Expanding $\delta \mathbf{u}(x,t)$ in exponentially growing (or decaying) Fourier modes $\delta \mathbf{u} = \mathbf{\hat{u}}_q \, e^{\sigma t} e^{iqx}$ leads to the eigenvalue problem
\begin{equation} \label{eq:eigenvalue-problem}
   \mathcal{J} \mathbf{\hat{u}}_q = \sigma \mathbf{\hat{u}}_q,
\end{equation}
with the Jacobian
\begin{equation*}
    \mathcal{J} = \begin{pmatrix}
    -D_c q^2 -i \vf q - f_c & -f_m \\
    f_c & -D_m q^2 + f_m
    \end{pmatrix},
\end{equation*}
where $f_c = \partial_c f |_{\mathbf{u}^*}$ and $f_m = \partial_m f |_{\mathbf{u}^*}$ encode the linearized reaction kinetics.
Note that for reaction kinetics of the form Eq.~\eqref{eq:att-det-kinetics}, $f_c = a(m) > 0$ and we consider this case in the following.

For each mode with wavenumber $q$, there are two eigenvalues $\sigma_{1,2}(q)$.
The case $q = 0$ corresponds to spatially homogeneous perturbations, where the two eigenvalues are given by $\sigma_1 = f_m - f_c$ and $\sigma_2 = 0$ \cite{Brauns.etal2018}.
Here, we restrict our analysis to homogeneously stable states ($\sigma_1 < 0$).
The second eigenvalue ($\sigma_2 = 0$) corresponds to perturbations that change the average mass $\bar{n}$ and therefore shift the homogeneous steady state $\mathbf{u}^*(\bar{n})$ along the nullcline $f = 0$. 
Because these perturbations break mass-conservation, they are not relevant for the stability of a closed system as considered here.
The modes $q > 0$ determine the stability of the system against spatially inhomogeneous perturbations (\emph{lateral stability}).
The eigenvalue with the larger real part determines the stability and will be denoted by $\sigma(q)$, suppressing the index.

A typical dispersion relation with a band of unstable modes is shown in Fig.~\ref{fig:2}A.
The real part (solid line), indicating the mode's growth rate, has a band of unstable modes $[0,q_\mathrm{max}]$ where $\Re \sigma(q) > 0$.
The fastest growing mode $q^*$ determines the wavelength $\lambda$ of the pattern that initially grows, triggered by a small, random perturbation of the spatially homogeneous steady state.
For $\vf = 0$, the imaginary part of $\sigma(q)$ vanishes, for locally stable steady states ($\sigma(0) \leq 0$). \cite{Brauns.etal2018}. 
However, in the presence of flow, the imaginary part of $\sigma(q)$ is non-zero (dashed line in Fig.~\ref{fig:2}A), which implies a propagation of each mode with the phase velocity $v_\mathrm{phase}(q) = - \Im \sigma(q) / q$.
This means that a mode $q$ not only grows over time (orange arrows in Fig.~\ref{fig:2}B), but also propagates as indicated by the pink arrows in Fig.~\ref{fig:2}B.
Further below, in Sec.~\ref{sec:slow-fast-flow-limit}, we will show that $\Im \sigma(q)$ always has the same sign as the flow velocity $\vf$, such that all modes propagate \emph{against} the flow direction. 

To gain physical insight into the mechanisms underlying the growth and propagation of perturbations (modes) we will first give an intuitive explanation of a lateral instability in McRD systems, building on the concepts of local equilibria theory \cite{Halatek.etal2018,Brauns.etal2018}.
We then provide a more detailed analysis in the limits of long wavelength as well as fast and slow flow.

\subsection{Intuition for the flow-driven instability and upstream propagation of the unstable mode} \label{sec:stability-intuition}

Lateral instability in McRD systems can be understood as a \emph{mass-redistribution instability} \cite{Brauns.etal2018}. Let us briefly recap the mechanism underlying this instability for a system without flow. 
To this end, we first discuss the effect of reactions and diffusion separately, and explain how these effects together drive the mass-redistribution instability.
We then explain how this instability is affected by cytosolic flow.

\begin{figure}
  \centering
  \includegraphics{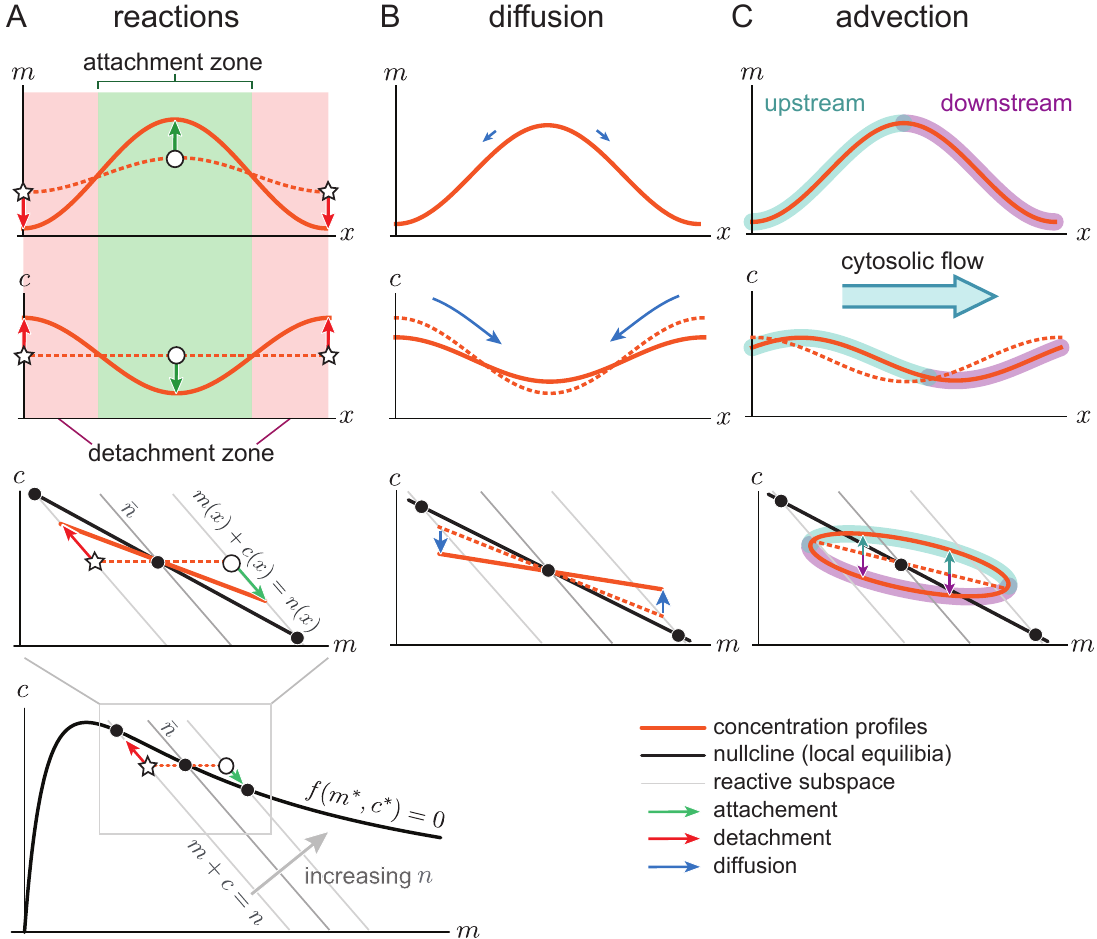}
  \caption{
    Sketch of the initial dynamics of an laterally unstable spatially homogeneous steady state. 
    The role of reactions (A), diffusion (B) and advection (C) for a mass-redistribution instability are presented for the membrane (top) and cytosolic (middle) concentration profiles and in phase space (bottom).
    (A) A small perturbation of the spatially homogeneous membrane concentration (orange dashed lines in top panel) leads to a spatially varying local total density $n(x)$, with a larger total density at the maximum of the membrane profile (open circle) and a smaller total density at the minimum (open star).
    These local variations in total density lead to attachment zones (green region) and detachment zones (red region). The reactive flow, indicated by the red and green arrows, points along the reactive subspace (gray lines) in phase space towards the shifted local equilibria (black circles). These reactive flows lead to the solid orange density profiles after a small amount of time.
    (B) Faster diffusion in the cytosol compared to the membrane (indicated by the large and small blue arrows in the middle and top panel, respectively), lead to net mass transport from the detachment zone to the attachment zone. Again, dashed and solid lines indicate the state before and after a short time interval of diffusive transport. 
    (C) Cytosolic flow shifts the cytosolic concentration with respect to the membrane concentration (orange dashed to orange solid lines), increasing the cytosolic concentration on the upstream side of the pattern and decreasing the cytosolic concentration on the downstream side . In phase space, the trajectory of this density profile forms a `loop'.
    }
\label{fig:3}
\end{figure}

Consider a spatially homogeneous steady state, perturbed by a slight redistribution of the \emph{local} total density $n(x,t)$.
The dashed orange line in Figure~\ref{fig:3}A shows such a perturbation where the membrane concentration (Fig.~\ref{fig:3}A top) is slightly perturbed in a sinusoidal fashion.
In phase space this is represented by a density distribution that slightly deviates from the spatially homogeneous steady state (marked by the orange dashed line). 
Here, the open star and open circle mark the minimum and maximum of the local total density, respectively.
The local total density determines the local reactive equilibrium concentrations $m^*(n)$ and $c^*(n)$ (cf.\ Eq.~\eqref{eq:HSS}, replacing the average mass $\bar{n}$ by the local mass $n(x,t)$). 
In phase space (Fig.~\ref{fig:3}A bottom) these local equilibria can be read off from the intersections (marked by black circles) of the reactive subspaces $n(x,t)= m(x,t)+c(x,t)$ (gray solid lines) and the reactive nullcline (black solid lines).
A slight redistribution of the local total density shifts the reactive equilibria, leading to reactive flows towards these shifted equilibria (red and green arrows in Fig.~\ref{fig:3}A).
Thus, the reactive equilibria, and thereby the reactive flows, are encoded in the shape of the reactive nullcline in phase space.
If the nullcline slope is negative, increasing the total density leads to a decreasing equilibrium cytosolic concentration and therefore to \emph{attachment} (green arrows in Fig.~\ref{fig:3}A).
Conversely, in regions of lower total density, the equilibrium cytosolic concentration increases via \emph{detachment} (red arrows in Fig.~\ref{fig:3}A).
Hence, regions of high total density become \emph{self-organized attachment zones} and regions of low total density become \emph{self-organized detachment zones}~\cite{Halatek.etal2018} (green and red areas in Fig.~\ref{fig:3} top and middle).

These attachment and detachment zones act as sinks and sources for diffusive mass-transport on the membrane and in the cytosol: 
The attachment zone acts as a cytosolic sink and membrane source, and the detachment zone acts as a cytosolic source and a membrane sink (blue arrows in Fig.~\ref{fig:3}B).
As diffusion in the cytosol is much faster than in the membrane, mass is transported faster in the cytosol than on the membrane, as indicated by the size of the blue arrows in Fig.~\ref{fig:3}B top and middle.
This leads to net mass transport from the detachment zone to the attachment zone.
As the local total density increases in the attachment zone, it facilitates further attachment and thereby the growth of the pattern on the membrane.
In short, the mechanism underlying the mass-redistribution instability is a cascade of attachment--detachment kinetics (Fig.~\ref{fig:3}A) and net mass-transport towards attachment zones (Fig.~\ref{fig:3}B).

How does cytosolic fluid flow affect the mass-redistribution instability?
Cytosolic flow transports proteins advectively.
This advective transport shifts the cytosolic density profile downstream relative to the membrane density profile (dashed to solid orange line in Fig.~\ref{fig:3}C middle).
This shift leads to an increase of the cytosolic density on the upstream (cyan) side of the membrane peak and a decrease on the downstream (magenta) side, in Fig.~\ref{fig:3}C (middle), respectively.
In phase space, this asymmetry is reflected as a `loop' shape of the phase space trajectory that corresponds to the real space pattern  (Fig.~\ref{fig:3}C bottom). 
The higher cytosolic density on the upstream side increases attachment relative to the downstream side.
This leads to a propagation of the membrane concentration profile in the upstream direction.

\subsection{Long wavelength limit} \label{sec:LSA-long-wavelength}
To complement this intuitive picture we consider the long wavelength limit $q \rightarrow 0$.%
\footnote{In principle, the dispersion relation can be easily obtained in closed form using the formula for eigenvalues of $2{\times}2$ matrices: $\sigma_{1,2} = \tfrac{1}{2}\tr \mathcal{J} \mp \tfrac{1}{2} \sqrt{(\tr \mathcal{J})^2 - 4 \det \mathcal{J}}$ where $\tr \mathcal{J}$ and $\det \mathcal{J}$ are the Jacobian's trace and determinant, respectively. Because the resulting expression is rather lengthy, we don't write it out it explicitly here.}
In this limit, the dispersion relation expanded to second order in $q$ reads 
\begin{equation} \label{eq:sigma-small-q}
    \sigma(q) \approx -\frac{1}{1 + \snc} \bigg[i \snc \vf q + (D_m + \snc D_c) q^2
       + \frac{\snc \, \vf^2}{f_c (1+\snc)^2}  q^2 \bigg],
\end{equation}
where $\snc = -f_m/f_c$ is the slope of the reactive nullcline.
The imaginary part $\Im \sigma(q)$ is linear in $q$ to lowest order, implying a phase velocity $v_\mathrm{phase} = \vf \snc /(1+\snc)$ that is independent of the wavelength.
The growth rate $\Re \sigma(q)$ is quadratic in $q$ to lowest order. 
If this quadratic term is positive, there is a band of unstable modes.\footnote{Homogeneous stability implies that the nullcline slope $\snc$ is larger than $-1$ \cite{Brauns.etal2018}, such that the prefactor $(1+\snc)^{-1}$ is positive.}
Hence, the criterion for a mass-redistribution instability can be expressed in terms of the nullcline slope \cite{Brauns.etal2018}
\begin{equation} 
\label{eq:slope-criterion}
    \snc < - \frac{D_m}{D_c} \left[1 + \frac{\vf^2}{(1+\snc)^2 D_c f_c}\right]^{-1} \,.
\end{equation}

In the absence of flow, $\vf = 0$, we recover the slope criterion $\snc < -D_m/D_c$ for a mass-redistribution instability driven by cytosolic diffusion~\cite{Brauns.etal2018}.
We find that flow always increases the range of instability since the second term in the square brackets monotonically increases with flow speed $|\vf|$.
Furthermore, the instability criterion becomes independent of the diffusion constants in the limit of fast flow ($|\vf| \gg \sqrt{D_c f_c}$).
The criterion for the (flow-driven) mass-redistribution instability then simply becomes $\snc < 0$, independently of the ratio of the diffusion constants.
This has the interesting consequence that, for sufficiently fast flow, a mass-redistribution instability can be driven solely via cytoplasmic flow, independent of diffusion.

\subsection{Limits of slow and fast flow} 
\label{sec:slow-fast-flow-limit}
To analyze the effect of flow for wavelengths away from the long wavelength limit it is instructive to consider the limit cases of slow and fast flow speed.

We first consider a limit where advective transport $(q\vf)^{-1}$ is slow compared either to the chemical reactions or to diffusive transport.
To lowest order in $\vf$, the dispersion relation is given by (see Appendix~\ref{appendix})
\begin{equation} \label{eq:sigma-slow-flow}
     \sigma(q) \approx \sigma^{(0)}(q) + i \frac{\vf q}{2} A(q),
\end{equation}
where the zeroth order term, $\sigma^{(0)}(q)$, is the dispersion relation in the absence of flow, which has no imaginary part~\cite{Brauns.etal2018} (cf.\ Eq.~\eqref{eq:sigma0}).
The function $A(q)$ is positive for all laterally unstable modes ($\Re \sigma(q) > 0$).
Equation~\eqref{eq:sigma-slow-flow} shows that to lowest order (linear in $\vf$) the effect of cytosolic flow is to induce propagation of the modes with the phase velocity $v_\mathrm{phase}(q) = \Im \sigma(q)/q \approx -\vf A(q)$. 
Since $A(q) > 0$ for laterally unstable modes, all growing perturbations propagate against the direction of the flow (as illustrated in Fig.~\ref{fig:2}B).

In the limit of fast flow (compared either to reactions or to cytosolic transport) we find that the dispersion relation (given by the eigenvalue problem Eq.~\eqref{eq:eigenvalue-problem}) reduces to
\begin{equation} \label{eq:sigma-fast-flow}
     \sigma(q) \approx f_m - D_m q^2 + i \frac{f_c f_m}{\vf q}
\end{equation}
for non-zero wavenumbers.
The real part of the dispersion relation in this fast flow limit becomes identical to the dispersion relation in the limit of fast diffusion~\cite{Brauns.etal2018}.
In both limits, cytosolic transport becomes (near) instantaneous.
In particular, in the limit of fast flow, advective transport completely dominates over diffusive transport in the cytosol such that the dispersion relation becomes independent of the cytosol diffusion constant $D_c$.

From the imaginary part of $\sigma(q)$, we obtain the phase velocity $v_\mathrm{phase} = -f_c f_m/(\vf q^2)$.
In other words, an increase in cytosolic flow leads to a \emph{decrease} of the phase velocity.
This is opposite to the slow flow limit discussed above, where the phase velocity increased linearly with the flow speed.

To rationalize these findings, we recall the propagation mechanism as discussed above.
There, we argued that a phase shift between the membrane and the cytosol pattern is responsible for the pattern propagation, as it leads to an asymmetry in the attachment--detachment balance upstream and downstream.
This phase shift increases with the flow velocity and eventually saturates at $\pi/4$.
\footnote{The phase shift can be read off from the real and imaginary parts of the eigenvectors in the linear stability analysis.}
On the other hand, the cytosol concentration gradients become shallower the faster the flow. To understand why this is, imagine a small volume element in the cytosol being advected with the flow. The faster the flow, the less time it has to interact with each point on the membrane it passes. Therefore, for faster advective flow, the attachment--detachment flux at the membrane is effectively diluted over a larger cytosolic volume.
This leads to a flattening of the cytosolic concentration profile (see Movie~2), and therefore a reduction in the upstream--downstream asymmetry of attachment.
As a result, in the limit of fast flow, the pattern propagates \emph{slower} the faster the flow, whereas, in the limit of slow flow, the pattern propagates \emph{faster} the faster the flow.
Thus, comparing these two limits, we learn that the phase velocity reaches a maximum at intermediate flow speeds.

\subsection{Summary and discussion of linear stability}

Let us briefly summarize our main findings from linear stability analysis.
We found that the leading order effect of cytosolic flow is to induce upstream propagation of patterns. 
This propagation is driven by the faster resupply of protein mass on the upstream side of the pattern compared to the downstream side.
A similar effect was previously found for vegetation patterns which move uphill because nutrients are transported downhill by water flow~\cite{Siero.etal2015}. 
Even though these systems are not strictly mass conserving, their pattern propagation underlies the same principle: 
The nutrient uptake in regions of high vegetation density creates a nutrient sink which is resupplied asymmetrically due to the downhill flow of water and nutrients.

Moreover, we used a phase-space analysis to explain how flow extends the range of parameters where where patterns emerge spontaneously, i.e.\ where the homogeneous steady state is laterally unstable.
This was previously shown mathematically for general two-component reaction--diffusion systems (not restricted to mass-conserving ones)~\cite{Siero.etal2015,Perumpanani.etal1995}.
Our analysis in the long wavelength limit explains the physical mechanism of this instability for mass-conserving systems: 
The flow-driven instability is a mass-redistribution instability, 
driven by a self-amplifying cascade of (flow-driven) mass transport and the self-organized formation of attachment and detachment zones (shifting reactive equilibria).
This shows that the instability mechanism is identical to the mass-redistribution instability that underlies pattern formation in systems without flow (i.e.\ where only diffusion drives mass transport)~\cite{Brauns.etal2018}.
For these systems, the instability strictly requires $D_c > D_m$.
In contrast, we find that for sufficiently fast flow, there can be a mass-redistribution instability even in the absence of cytosolic diffusion ($D_c = 0$).
While the case $D_c = 0$ is not physiologically relevant in the context of intracellular pattern formation, it may be relevant for the formation of vegetation patterns on sloped terrain~\cite{Samuelson.etal2019}, where $c$ and $m$ are the soil-nutrient concentration and plant biomass density, respectively.
In conclusion, advective flow can fully replace diffusion as the mass-transport mechanism driving the mass-redistribution instability.

\section{Pattern propagation in the nonlinear regime}
\label{sec:steady-state}

\begin{figure}
  \centering
  \includegraphics{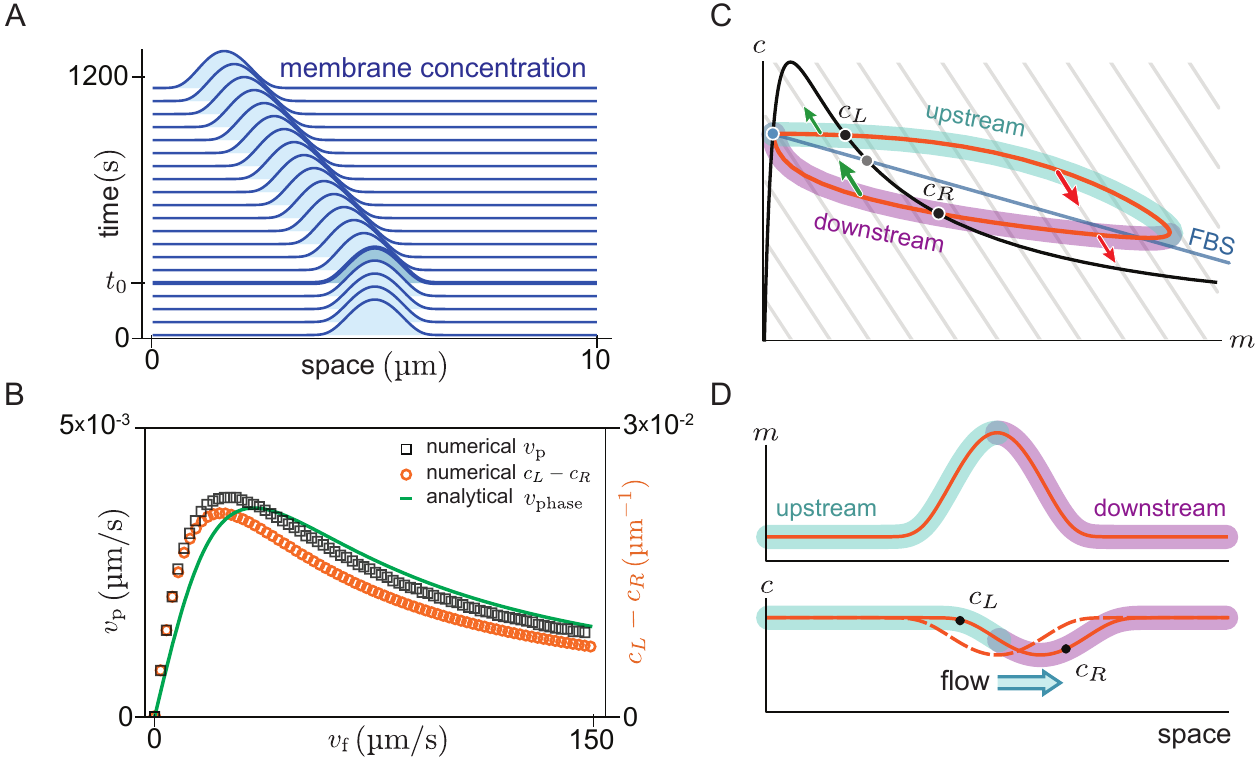}
  \caption{
  Pattern dynamics far from the spatially homogeneous steady state.
  (A) Time evolution of the membrane-bound protein concentration. At time $t_0 = \SI{240}{s}$ a constant cytosolic flow with velocity $\vf=\SI{20}{\micro m/s}$ towards the right is switched on (cf.\ Movie~3).
  (B) Relation between the peak speed ($\vp$) and flow speed ($\vf$). Results from finite element simulations (black open squares) are compared to the phase velocity of the mode $q_\mathrm{max}$ obtained from linear stability analysis (green solid line) and to an approximation (orange open circles) of the area enclosed by the density distribution trajectory in phase space (area enclosed by the `loop' in D).
  (The domain size, $L = \SI{10}{\micro m}$, is chosen large enough compared to the peak width such that boundary effects are negligible.)
  (C) A schematic of the phase portrait corresponding to the pattern in D. The density distribution in the absence of flow is embedded in the FBS (blue straight line).
  In the presence of flow, the density distribution trajectory forms a `loop' in phase space. The upstream and downstream side of the pattern are highlighted in cyan and magenta, respectively.
  Red and green arrows indicate the direction of the reactive flow in the attachment and detachment zones, respectively.
  At intersection points of the density distribution with the nullcline ($c_\mathrm{L}$ and $c_\mathrm{R}$) the system is at its local reactive equilibrium. 
  (D) Sketch of the membrane (orange solid line, top) and cytosolic (orange dashed line, bottom) concentration profiles for a stationary pattern in the absence of cytosolic flow.
  Flow shifts the cytosol profile downstream (orange solid line, bottom).
  }
  \label{fig:4}
\end{figure}

So far we have analyzed how cytosolic flow affects the dynamics of the system in the vicinity of a homogeneous steady state, using  linear stability analysis. 
However, patterns generically don't saturate at small amplitudes but continue to grow into the strongly nonlinear regime~\cite{Brauns.etal2018} (see Movie~1 for an example in which a small perturbation of the homogeneous steady state evolves into a large amplitude pattern in the presence of flow).

To study the long time behavior (steady state) far away from the spatially homogeneous steady state, we performed finite element simulations in Mathematica~\cite{Mathematica}.
To interpret the results of these numerical simulations, we will use \emph{local equilibria theory}, building on the phase-space analysis introduced in Refs.~\cite{Brauns.etal2018, Halatek.etal2018}.

Figure \ref{fig:4}A shows the space-time plot (kymograph) of a system where there is initially no flow ($t < t_0$), such that the system is in a stationary state with a single peak.
For such a stationary steady state, diffusive fluxes on the membrane and in the cytosol have to balance exactly. 
This diffusive flux balance imposes the constraint that in the $(m,c)$-phase plane, the trajectory corresponding to the pattern lies on a straight line with slope $-D_m/D_c$, called `\emph{flux-balance subspace}' (FBS)~\cite{Brauns.etal2018} (see light blue line in Fig.~\ref{fig:4}C).
At the plateaus of the pattern, diffusive flow vanishes and attachment and detachment are balanced, i.e.\ the system is locally in reactive equilibrium. 
Hence, plateaus corresponds to points in the $(m,c)$-phase plane where the FBS intersects the reactive nullcline on a segment with slope larger than $-D_m/D_c$ (blue point in Fig.~\ref{fig:4}C).
The intersection point between the FBS and the nullcline where the nullcline slope is smaller than $-D_m/D_c$ (gray point in Fig.~\ref{fig:4}C) corresponds inflection points of the pattern profile. 
An in depth analysis of stationary patterns based on these geometric relations in phase space can be found in Ref.~\cite{Brauns.etal2018}.
Here we ask how the phase portrait changes in the presence of flow.

At time $t = t_0$, a constant cytosolic flow in the positive $x$-direction is switched on.
Consistent with the expectation from linear stability analysis, we find that the peak propagates against the flow direction in the negative $x$-direction (solid lines in Fig.~\ref{fig:4}A). 
The diffusive fluxes no longer balance for this propagating steady state, such that the phase-space trajectory is no longer embedded in the FBS.
Instead, as advective flow shifts the cytosol concentration profile relative to the membrane profile, the phase-space trajectory becomes a `loop' (Fig.~\ref{fig:4}C).
On the upstream side of the peak, the cytosolic density is increased, such that net attachment --- which is proportional to the cytosolic density --- is increased relative to net detachment. Conversely, the reactive balance is shifted towards detachment on the downstream side. 
Because the reactive flow is approximately proportional to the distance from the reactive nullcline in phase space, the asymmetry between net attachment and detachment on the upstream and downstream side of the peak can be estimated by the area enclosed by the loop-shaped trajectory in phase space.

To test whether the attachment--detachment asymmetry explains the propagation speed of the peak, we estimate the enclosed area in phase space by the difference in cytosolic concentrations at the points $c_\mathrm{L}$ and $c_\mathrm{R}$ (black dots in Fig.~\ref{fig:4}C and D) where the loop intersects the reactive nullcline ($f = 0$ black line Fig.~\ref{fig:4}C). At these points, the system is in a local reactive equilibrium.
Indeed, we find that the propagation speed of the pattern obtained from numerical simulations (black open squares in Fig.~\ref{fig:4}B) is well approximated by the difference in cytosolic density ($\vp \propto c_\mathrm{L} - c_\mathrm{R} $) for all flow speeds (orange open circles in Fig.~\ref{fig:4}B).
Furthermore, in the limit of slow and fast flow, the peak propagation speed is well approximated by the propagation speed of the unstable traveling mode with the longest wavelength, as obtained from linear stability analysis.%
\footnote{The phase velocity depends on the mode's wavelength. The relevant length scale for the peak's propagation is its width, which is approximately given by $2\pi/q_\mathrm{max}$ at the pattern's inflection point~\cite{Brauns.etal2018}. Thus, we infer the peak propagation speed from $\Im\sigma(q_\text{max})/q_\mathrm{max}$ at the inflection point of the stationary peak.}
For small flow speeds, the pattern's propagation speed $\vp$ increases linearly with $\vf$ (cf. Eq.~\ref{eq:sigma-small-q}) and for large flow speeds the pattern speed is proportional to $1/\vf$ (cf. Eq.~\ref{eq:sigma-fast-flow}).

In summary, we found that the peak propagation speed in the slow and fast flow limits is well described by the propagation speed of the linearly unstable mode with the longest wavelength (i.e.\ the right edge of the band of unstable modes $q_\mathrm{max}$).
Moreover, we approximated the asymmetry of protein attachment by the area enclosed by the density distribution in phase space, and found that this is proportional to the peak speed for all flow speeds.

\section{Flow-induced transition from mesa to peak patterns}
\label{sec:mesa-peak}

\begin{figure}
  \centering
  \includegraphics[width=\linewidth]{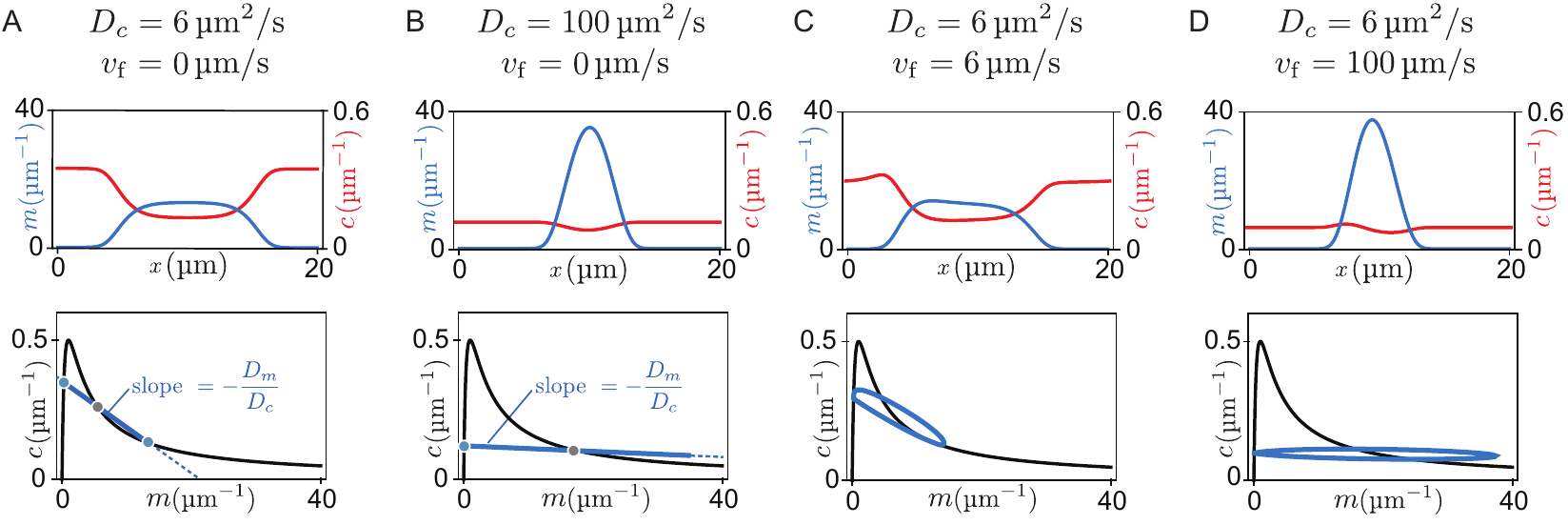}
  \caption{
  Demonstration of the transition from a mesa pattern to a peak pattern. Each panel shows a snapshot from finite element simulations in steady state. Top concentration profiles in real space; bottom: corresponding trajectory (blue solid line) in phase space.
  (A) Mesa pattern in the case of slow cytosol diffusion and no flow. The two plateaus (blue dots) and the inflection point (gray dot) of the pattern correspond to the intersection points of the FBS (blue dashed line) with the reactive nullcline (black line).
  (B) For fast cytosol diffusion, the third intersection point between FBS and nullcline lies at much higher membrane concentration such that it no longer limits the pattern amplitude. Therefore, a peak forms whose amplitude is limited by the total protein mass in the system.
  (C) Slow flow only slightly deforms the mesa pattern, compare to (A).
  Fast cytosolic flow leads to formation of a peak pattern (D), similarly to fast diffusion. Parameters: $\bar{n}=\SI{7}{\micro m^{-1}}, D_m=\SI{0.1}{\micro m^2/s}$ and $L=\SI{20}{\micro m}$.
  }
  \label{fig:5}
\end{figure}

So far we have studied the propagation of patterns in response to cytosolic flow.
Next, we will show how cytosolic flow can also drive the transition between qualitatively different pattern types.
We distinguish two pattern types exhibited by McRD systems, peaks and mesas~\cite{Chiou.etal2018,Brauns.etal2018}.
Mesa patterns are composed of plateaus (low density and high density) connected by interfaces, while a peak can be pictured as two interfaces concatenated directly (cf.\ Fig.~\ref{fig:4}).
Mesa patterns form if protein attachment saturates in regions of high total density, while peaks form if the attachment rate does not saturate at high density~\cite{Brauns.etal2018,Chiou.etal2018}. 
Thus, while the amplitude of mesa patterns is determined by the attachment--detachment balance in the two plateaus, the amplitude (maximum concentration) of a peak is determined by the total mass available in the system~\cite{Brauns.etal2018}.

How does protein transport affect whether a peak or a mesa forms?
As we argued above, a peak pattern forms if protein attachment in regions of high density does not saturate.
In general, this will happen if attachment to the membrane depletes proteins from the cytosol slower than lateral transport can resupply proteins (see Fig.~\ref{fig:5}A).
Let us first recap the situation without flow, where proteins are resupplied by diffusion from the detachment zone to the attachment zone across the pattern's interface with width $\ell_\mathrm{int}$.
Thus, a peak pattern forms if the rate of transport by cytosolic diffusion is faster than the attachment rate ($D_c/\ell_\mathrm{int}^2 \gg \tau_\mathrm{react}^{-1}$).
Further using that the interface width is given by a balance of membrane diffusion and local reactions ($\ell_\mathrm{int}^2 \sim \tau_\mathrm{react} D_m$),
we obtain the condition $D_c \gg D_m$ for the formation of peak patterns.

In terms of phase space geometry, this means that the slope $-D_m/D_c$ of the flux-balance subspace in phase space must be sufficiently shallow.
For a steep slope $-D_m/D_c$ of the FBS, the membrane concentration saturates at the point where the FBS intersects with the reactive nullcline blue dots in Fig.~\ref{fig:5}A. There, attachment and detachment balance such that a mesa forms (Fig.~\ref{fig:5}A).
For faster cytosol diffusion, the flux-balance subspace is shallower such that the third FBS-NC intersection point shifts to higher densities. Thus, for sufficiently fast cytosol diffusion a peak forms (Fig.~\ref{fig:5}B).

Adding slow cytosolic flow does not significantly contribute to the resupply of the cytosolic sink (i.e.\ attachment zone) and therefore does not alter the pattern type
(Fig.~\ref{fig:5}C). 
In contrast, when cytosolic protein transport (by advection and/or diffusion) is fast compared to the reaction kinetics, the cytosolic sink gets resupplied quickly, leading to a flattening of the cytosolic concentration profile.
Accordingly, the density distribution in phase space approaches a horizontal line, both for fast cytosolic diffusion
(Fig.~\ref{fig:5}B) and
for fast cytosolic flow (Fig.~\ref{fig:5}D).
As a consequence, the point where the density distribution meets the nullcline shifts towards larger membrane concentrations, resulting in an increasing amplitude of the mesa pattern.
Eventually, when the amplitude of the pattern can not grow any further due to limiting total mass, a peak pattern forms (Fig.~\ref{fig:5}B,D). 
Hence, an increased flow velocity can cause a transition from a mesa pattern to a peak pattern (see Movie~4).

In summary, we found that cytosolic flow can qualitatively change the membrane-bound protein pattern from a small-amplitude, wide mesa pattern to a large-amplitude, narrow peak pattern.
In cells, such flows could therefore promote the precise positioning of polarity patterns on the membrane.
Furthermore, we hypothesize that flow can contribute to the selection of a single peak by accelerating the coarsening dynamics of the pattern via two distinct mechanisms.
First, flow accelerates protein transport that drives coarsening.
Second, as peak patterns coarsen faster than mesa patterns~\cite{Chiou.etal2018,Brauns.etal2020}, flow can accelerate coarsening via the flow-driven mesa-to-peak transition.
Such fast coarsening may be important for the selection of a single polarity axis, e.g.\ a single budding site in \textit{S.~cerevisiae} \cite{Chiou.etal2017}, for axon formation in neurons \cite{Fivaz.etal2008}, and to establish a distinct front and back in motile cells \cite{Wang.etal2013,Keilberg.Sogaard-Andersen2014}.

\section{Flow-induced pattern formation}
\label{sec:pattern-trigger}

\begin{figure}
  \centering
  \includegraphics[width=\linewidth]{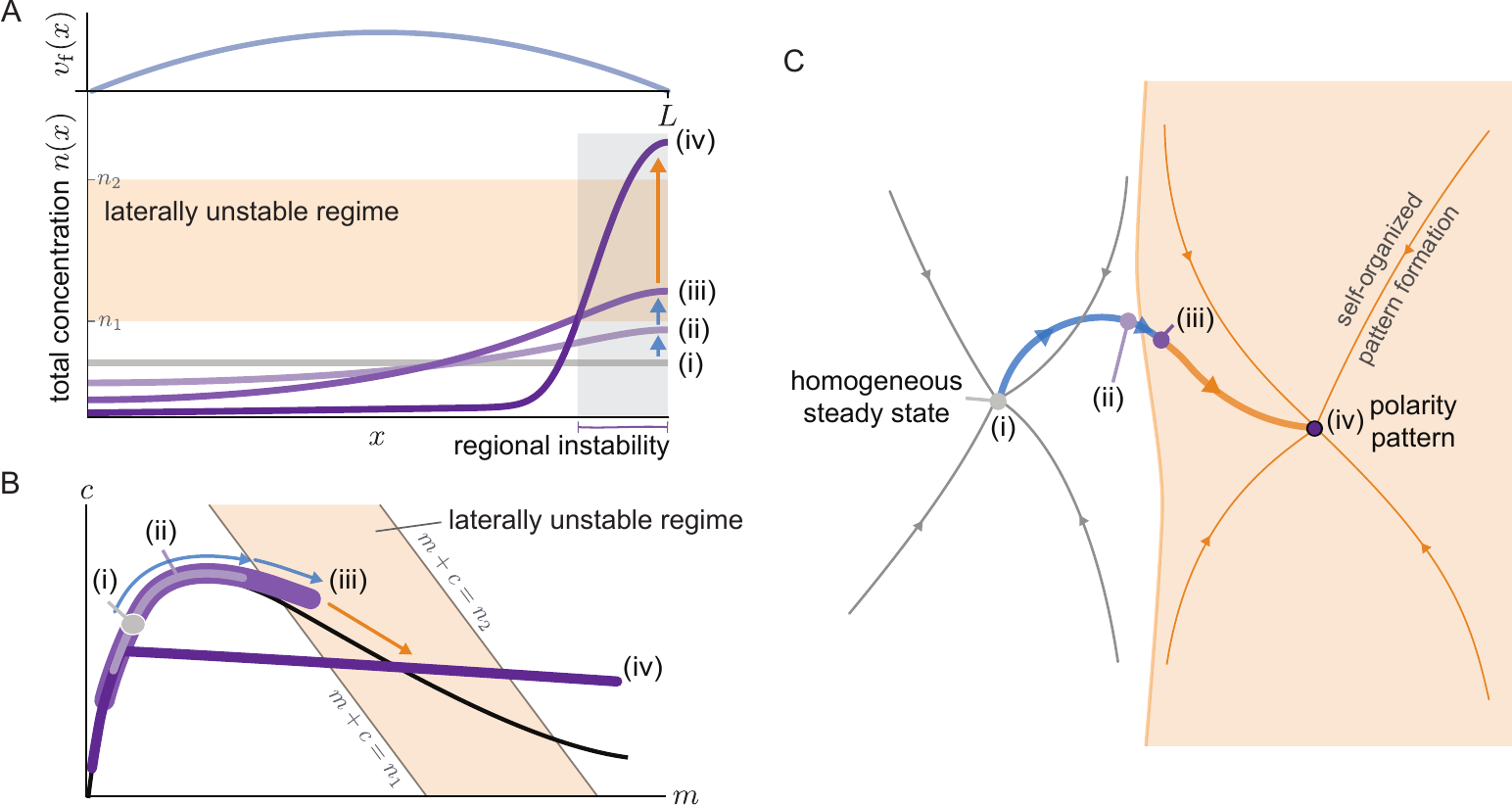}
  \caption{
  Flow-driven protein mass accumulation can induce pattern formation by triggering a regional lateral instability.
  (A) Top: quadratic flow velocity profile: $v_f = -v_\mathrm{max} \, \left(x/L - 1/2\right)^2$. Bottom: illustration of the total density profiles at different time points starting from a homogeneous steady state (\textit{i}) to the final pattern (\textit{iv}); see Movie~5. Mass redistribution due to the non-uniform flow velocity drives mass towards the right hand side of the system, as indicated by the blue arrows. The range of total densities shaded in orange indicates the laterally unstable regime determined by linear stability analysis.
  Once the total density reaches this regime locally, a regional lateral instability is triggered resulting in the self-organized formation of a peak (orange arrow).
  (B) Sketch of the phase space representation corresponding to the profiles shown in A. Note that the concentrations are slaved to the reactive nullcline (black line) until the regional lateral instability is triggered.
  (C) Schematic representation of the state space of concentration patterns in a case where both the homogeneous steady state and a stationary polarity pattern are stable. Thin trajectories indicate the dynamics in the absence of flow and the pattern's basin of attraction is shaded in orange. The thick trajectory connecting both steady states shows the flow-induced dynamics, corresponding to the sequence of states (i)--(iv) shown in A and B.
  }
  \label{fig:6}
\end{figure}

So far we have studied how a uniform flow profile affects pattern formation on a domain with periodic boundary conditions, representing circular flows along the cell membrane and bulk flows in microfluidic \textit{in vitro} setups.

However, flows in the vicinity of the membrane can be non-uniform. We will discuss examples of such non-uniform flows at the end of this section.
A non-uniform flow transports the proteins at different speeds along the membrane. Starting from a spatially homogeneous initial state, this leads to a redistribution of mass. 
It has been demonstrated in previous work that this non-uniform flow can induce pattern formation even if the homogeneous steady state is laterally stable (i.e.\ there is no spontaneous pattern formation)~\cite{Goehring.etal2011, Diegmiller.etal2018, Gross.etal2019}.
Based on numerical simulations, a transition from flow-guided to self-organized dynamics has been reported~\cite{Gross.etal2019}.
However, the physical mechanism underlying this transition, and what determines the transition point have remained unclear.

As a minimal system to address this question, we consider a one-dimensional domain with no-flux boundaries and a parabolic speed profile that vanishes at the system boundaries (Fig.~\ref{fig:6}A, top).
In the following, we describe the flow-induced dynamics starting from a spatially homogeneous steady state to the final polarity pattern observed in numerical simulations (see Movie~5). Figure~\ref{fig:6} visualizes these dynamics in real space (A) and in the $(m,c)$-phase plane (B).
To relate our findings to the previous study Ref.~\cite{Gross.etal2019}, we also visualize the dynamics in an abstract representation of the state space (comprising all concentration profiles) used in this previous study. In this state space, steady states are points and the time evolution of the system is a trajectory (thick blue/orange line in Fig.~\ref{fig:6}C).

Starting from the homogeneous steady state (\textit{i}), the non-uniform advective flow redistributes mass in the cytosol (\textit{ii}).
Due to this redistribution of mass, the local reactive equilibria shift as we have seen repeatedly here and in earlier studies of mass-conserving systems~\cite{Halatek.Frey2018,Brauns.etal2018}.
In fact, as long as the gradients of both the membrane and cytosol profiles are shallow, the concentrations remain close to the local equilibria, as evidenced by the density distribution in phase space spreading along the reactive nullcline (see profile (ii) in Fig.~\ref{fig:6}A,B). 
Eventually, the region where mass accumulates (here the right edge of the domain),
enters the laterally unstable regime (see profile \textit{iii}).
In the phase plane (Fig.~\ref{fig:6}B) this regime corresponds to the range of total densities $\bar{n}$ where the nullcline slope has a steeper negative than the flux-balance subspace ($\snc < -D_m/D_c$)%
\footnote{More precisely, the size of the laterally unstable region must be larger than the shortest unstable mode (corresponding to the right edge of the band of unstable modes in the dispersion relation (Fig.~\ref{fig:2}A)).
}. 
The mass-redistribution instability in this region, based on the self-organized formation of attachment and detachment zones (cf.\ Sec.~\ref{sec:stability-intuition}) will lead to the formation of a polarity pattern there (\textit{iv}).
Thus, the onset of a \emph{regional lateral instability} marks the transition from flow-guided dynamics to self-organized dynamics.

In the abstract state space visualization (Fig.~\ref{fig:6}C) the area shaded in orange indicates the polarity pattern's basin of attraction comprising all states (concentration profiles) where a spatial region in the system is laterally unstable.
In the absence of flow, states that do not exhibit such a laterally unstable region return to the homogeneous steady state (thin gray lines).
Non-uniform cytosolic flow induces mass-redistribution, that can drive an initially homogeneous system (i) into the polarity pattern's basin of attraction. From there on, self-organized pattern formation takes over, leading to the formation of a polarity pattern (iv), essentially independently of the advective flow (orange trajectory).

Similar pattern forming mechanisms based on a regional instability have previously been shown to also underlie stimulus-induced pattern formation following a sufficiently strong initial perturbation~\cite{Brauns.etal2018} and peak formation at a domain edge where the reaction kinetics abruptly change~\cite{Wigbers.etal2020}.
Thus, an overarching principle for stimulus-induced pattern formation emerges: To trigger (polarity) pattern formation, the stimulus, be it advective flow or heterogeneous reaction kinetics, has to redistribute protein mass in a way such that a regional (lateral) instability is triggered.

It remains to be discussed what happens once the cytoplasmic flow is switched off after the polarity pattern has formed.
In general, the polarity pattern will persist (see Movie~5), since it is maintained by self-organized attachment and detachment zones, largely independent of the flow.
However, as long as there is flow, the average mass on the right hand side of the system (downstream of the flow) is higher than on the left hand side.
Hence, flow can maintain a polarity pattern even if the average mass in the system as a whole is too low to sustain polarity patterns in the absence of flow (see bifurcation analysis in Ref.~\cite{Brauns.etal2018}).
If this is the case, the peak disappears once the flow is switched off (see Movie~6).

In summary, the redistribution of the protein mass is key to induce (polarity) pattern formation starting from a stable homogeneous state.
There are different scenarios how intracellular flows can lead to such mass redistribution:
First, one (or more) components of the pattern forming system may be embedded in the cell cortex \cite{Goehring.etal2011, Wang.etal2017, Gross.etal2019} which is a contractile medium driven by myosin-motor activity.
Indeed, it was previously demonstrated that advection of proteins in the cell cortex can induce a polarity pattern in a conceptual 2cMcRD model~\cite{Diegmiller.etal2018} and more quantitative models for the PAR-system~\cite{Goehring.etal2011, Gross.etal2019}.
Second, three-dimensional flows in the cytoplasm can result in local accumulation of protein mass on the membrane due to flow in the direction normal to the membrane.
Thus, the 3D flow field of the cytosol, which is incompressible, can have a similar effect as compressible cortex flows~\cite{Mittasch.etal2018}.

\section{Conclusions and outlook}

Inside cells, proteins are transported via diffusion and fluid flows, which, in combination with reactions, can lead to the formation of protein patterns on the cell membrane. 
To characterize the role fluid flows play in pattern formation, we studied the effect of flow on the formation of a polarity pattern, using a generic two-component model. 
We found that flow leads to propagation of the polarity pattern \emph{against} the flow direction with a speed that is maximal for intermediate flow speeds, i.e.\ when the rate of advective transport is comparable to either the reaction rates or to the rate diffusive transport in the cytosol.
Using a phase-space analysis, we showed that the propagation of the pattern is driven by an asymmetric influx of protein mass to a self-organized protein-attachment zone. As a consequence, attachment is stronger on the upstream side of the pattern compared to the downstream side, leading to upstream propagation of the membrane bound pattern.
Furthermore, we have shown that flow can qualitatively change the pattern from a wide mesa pattern (connecting two plateaus) to a narrow peak pattern. 
Finally, we have presented a phase-space analysis to elucidate the interplay between flow-guided dynamics and self-organized pattern formation.
This interplay was previously studied numerically in the context of PAR-protein polarization~\cite{Goehring.etal2011,Gross.etal2019}.
Our analysis reveals the underlying cause for the transition from flow-guided to self-organized dynamics: the regional onset of a mass-redistribution instability.

We discussed implications of our results and links to earlier literature at the end of each section.
Here, we conclude with a brief outlook.
We expect that the insights obtained from the minimal two-component model studied here generalize to systems with more components and multiple protein species.
For example, \emph{in vitro} studies of the reconstituted MinDE system of \emph{E.~coli} show that MinD and MinE spontaneously form dynamic membrane-bound patterns, including spiral waves~\cite{Loose.etal2008} and quasi-stationary patterns~\cite{Glock.etal2019}.
These patterns emerge from the competition of MinD self-recruitment and MinE-mediated detachment of MinD~\cite{Huang.etal2003,Halatek.Frey2012}.
In the presence of a bulk flow, the traveling waves were found to propagate upstream~\cite{Ivanov.Mizuuchi2010}.
Our analysis based on a simple conceptual model suggests that this upstream propagation is caused by a larger influx of the self-recruiting MinD on the upstream flanks compared to the downstream flanks of the travelling waves.
However, the bulk flow also increases the resupply of MinE on the upstream flanks. As MinE mediates the detachment of MinD and therefore effectively antagonizes MinD's self-recruitment, this may drive the membrane-bound patterns to propagate downstream instead of upstream. 
Which one of the two processes dominates --- MinD-induced upstream propagation or MinE-induced downstream propagation --- likely depends on the details of their interactions. 
This interplay will be the subject of future work.

A different route of generalization is to consider advective flows that depend on the protein concentrations. 
In cells, such coupling arises, for instance, from myosin-driven cortex contractions \cite{Gross.etal2019,Deneke.etal2019} and shape deformations \cite{Koslover.etal2017,Klughammer.etal2018}.
Myosin-motors, in turn, may be advected by the flow and their activity is controlled by signalling proteins such as \mbox{GTP}ases and kinases \cite{Iden.Collard2008}.
This can give rise to feedback loops between flow and protein patterns.
Previous studies show that such feedback loops can give rise to mechano-chemical instabilities \cite{Bois.etal2011}, drive pulsatile (standing-wave) patterns \cite{Radszuweit.etal2013,Kumar.etal2014} or cause the breakup of traveling waves \cite{Nakagaki.etal1999}.
We expect that our analysis based on phase-space geometry can provide insight into the mechanisms underlying these phenomena.

\vspace{6pt} 


\authorcontributions{All authors designed and carried out the research; MCW, FB and EF wrote the paper; CYL visualized the findings.}

\funding{This work was funded by the Deutsche Forschungsgemeinschaft (DFG, German Research Foundation) -- Project-ID 201269156 --  Collaborative Research Center (SFB) 1032 -- Project B2. M.C.W. and C.Y.L. are supported by a DFG fellowship through the Graduate School of Quantitative Biosciences Munich (QBM). M.C.W. acknowledges the Joachim Herz Stiftung for support.}

\conflictsofinterest{The authors declare no conflict of interest.} 

\abbreviations{The following abbreviations are used in this manuscript:\\

\noindent 
\begin{tabular}{@{}ll}
McRD & mass-conserving reaction--diffusion\\
2cMcRD & two-component mass-conserving reaction--diffusion\\
FBS & flux-balance subspace\\
\end{tabular}}

\appendixtitles{yes} 
\appendix
\section{Limit of slow flow and timescale comparison} \label{appendix}
The dispersion relation in the absence of flow ($\vf = 0$) reads
\begin{equation}
\label{eq:sigma0}
    \sigma^{(0)}=-\frac{1}{2}\left[\left(D_m+D_c\right)q^2+f_c-f_m\right]+\frac{B(q)}{2A(q)},
\end{equation}
with $A(q) = \big[1- 4f_c f_m/B(q)^2\big]^{-1/2} - 1$ and $B(q) = f_m + f_c + (D_c - D_m)q^2$.
To find the effect of slow flow, we first need to identify the relevant timescales such that we can define a dimensionless small parameter to expand in.
Because pattern formation is driven by transport in the cytosol (diffusive and advective) and attachment from the cytosol to the membrane, there are three relevant timescales: (\textit{i}) The the rate of advective transport on length scale $q^{-1}$ is given by $q \vf$; (\textit{ii}) The rate of diffusive transport on that scale, given by $D_c q^2$; and (\textit{iii}) the attachment rate $f_c = a(m)$ (cf.\ Eq.~\eqref{eq:att-det-kinetics}).
To compare these timescales, we form two dimensionless numbers:
the Pecl\'et number $\mathrm{Pe} = \vf/(D_c q)$ and
the Damk\"ohler number $\mathrm{Da} = f_c/(\vf q)$.
Flow can either be slow compared to reactions ($\mathrm{Da} \gg 1$) or slow compared to diffusion ($\mathrm{Pe} \ll 1$).
In both cases, expanding the the dispersion relation $\sigma(q)$ to first order yields
\begin{equation}
     \sigma(q) = \sigma^{(0)}(q) + i \frac{\vf q}{2} A(q) + \mathcal{O}(\varepsilon^2),
\end{equation}
where $\varepsilon = \min(\mathrm{Pe},\mathrm{Da}^{-1})$. 
By elementary algebra using the assumptions $D_c > D_m$ and $f_c > 0$ made above, it follows that $A(q)$ is positive when $\snc < 0$.
As Eq.~\eqref{eq:slope-criterion} in Sec.~\ref{sec:LSA-long-wavelength} shows, the condition $\snc < 0$ is necessary for a band of unstable modes to exist.
Therefore, $A(q)$ is positive for all unstable modes.

\section{Movie descriptions}
\begin{enumerate}
    \item Growth of a pattern from a     homogeneous steady state in the presence of flow. Top: concentration profiles in space; bottom: corresponding density distribution in the phase space. (Parameters: $D_m=\SI{0.1}{\micro m^2 /s}, \bar{n}=\SI{3}{\micro m^{-1}}$, $L=\SI{20}{\micro m}$ and $\vf=\SI{20}{\micro m /s}$.) 
    \item Simulation with adiabatically increasing flow speed from $\vf=\SI{0}{\micro m /s}$ to $\vf=\SI{100}{\micro m /s}$. Note the flattening of the cytosolic concentration profile as the flow speed increases. (Fixed parameters as for Movie~1.)
    \item Corresponds to the space-time plot in Fig.~\ref{fig:4}A.
    \item Pattern transformation from a mesa pattern to a peak pattern as flow speed is adiabatically increased from $\vf=0$ to $\vf=\SI{45}{\micro m /s}$. (Fixed parameters as for Movie~1.)
    \item Pattern formation triggered by mass-redistribution due to a spatially non-uniform flow (parabolic flow profile shown in Fig.~\ref{fig:6}A). After the flow is switched off at $t =\SI{200}{s}$, the pattern is maintained. (Parameters: $D_m=\SI{0.1}{\micro m^2/s}, L=\SI{30}{\micro m}$, $v_\mathrm{max} =\SI{1}{\micro m /s}$, and $\bar{n}=\SI{1}{\micro m^{-1}}$.)
    \item As Movie~5, but with lower average mass, $\bar{n}=\SI{0.8}{\micro m^{-1}}$. This mass is not sufficient to maintain a stationary peak in the absence of flow. Therefore, the peak disappears after the flow is switched off ($t > \SI{200}{s}$).
\end{enumerate}

\reftitle{References}

\externalbibliography{yes}
\bibliography{cytoplasmic_flow}

\begin{thebibliography}{-------}
\providecommand{\natexlab}[1]{#1}

\bibitem[Weiner(2002)]{Weiner2002}
Weiner, O.D.
\newblock Regulation of Cell Polarity during Eukaryotic Chemotaxis: The
  Chemotactic Compass.
\newblock {\em Current Opinion in Cell Biology} {\bf 2002}, {\em 14},~196--202.
\newblock
  doi:{\changeurlcolor{black}\href{https://doi.org/10.1016/S0955-0674(02)00310-1}{\detokenize{10.1016/S0955-0674(02)00310-1}}}.

\bibitem[Keilberg and
  {S{\o}gaard-Andersen}(2014)]{Keilberg.Sogaard-Andersen2014}
Keilberg, D.; {S{\o}gaard-Andersen}, L.
\newblock Regulation of {{Bacterial Cell Polarity}} by {{Small GTPases}}.
\newblock {\em Biochemistry} {\bf 2014}, {\em 53},~1899--1907.
\newblock
  doi:{\changeurlcolor{black}\href{https://doi.org/10.1021/bi500141f}{\detokenize{10.1021/bi500141f}}}.

\bibitem[Bi and Park(2012)]{Bi.Park2012}
Bi, E.; Park, H.O.
\newblock Cell {{Polarization}} and {{Cytokinesis}} in {{Budding Yeast}}.
\newblock {\em Genetics} {\bf 2012}, {\em 191},~347--387.
\newblock
  doi:{\changeurlcolor{black}\href{https://doi.org/10.1534/genetics.111.132886}{\detokenize{10.1534/genetics.111.132886}}}.

\bibitem[Chiou \em{et~al.}(2017)Chiou, Balasubramanian, and
  Lew]{Chiou.etal2017}
Chiou, J.g.; Balasubramanian, M.K.; Lew, D.J.
\newblock Cell {{Polarity}} in {{Yeast}}.
\newblock {\em Annual Review of Cell and Developmental Biology} {\bf 2017},
  {\em 33},~77--101.
\newblock
  doi:{\changeurlcolor{black}\href{https://doi.org/10.1146/annurev-cellbio-100616-060856}{\detokenize{10.1146/annurev-cellbio-100616-060856}}}.

\bibitem[Goryachev and Leda(2017)]{Goryachev.Leda2017}
Goryachev, A.B.; Leda, M.
\newblock Many Roads to Symmetry Breaking: Molecular Mechanisms and Theoretical
  Models of Yeast Cell Polarity.
\newblock {\em Molecular Biology of the Cell} {\bf 2017}, {\em 28},~370--380.
\newblock
  doi:{\changeurlcolor{black}\href{https://doi.org/10.1091/mbc.e16-10-0739}{\detokenize{10.1091/mbc.e16-10-0739}}}.

\bibitem[Turing(1952)]{Turing1952}
Turing, A.M.
\newblock The Chemical Basis of Morphogenesis.
\newblock {\em Philosophical Transactions of the Royal Society of London.
  Series B, Biological Sciences} {\bf 1952}, {\em 237},~37--72.
\newblock
  doi:{\changeurlcolor{black}\href{https://doi.org/10.1098/rstb.1952.0012}{\detokenize{10.1098/rstb.1952.0012}}}.

\bibitem[Hawkins \em{et~al.}(2009)Hawkins, B{\'e}nichou, Piel, and
  Voituriez]{Hawkins.etal2009}
Hawkins, R.J.; B{\'e}nichou, O.; Piel, M.; Voituriez, R.
\newblock Rebuilding Cytoskeleton Roads: {{Active}}-Transport-Induced
  Polarization of Cells.
\newblock {\em Physical Review E} {\bf 2009}, {\em 80},~040903.
\newblock
  doi:{\changeurlcolor{black}\href{https://doi.org/10.1103/PhysRevE.80.040903}{\detokenize{10.1103/PhysRevE.80.040903}}}.

\bibitem[Calvez \em{et~al.}(2020)Calvez, Lepoutre, Meunier, and
  Muller]{Calvez.etal2020}
Calvez, V.; Lepoutre, T.; Meunier, N.; Muller, N.
\newblock Non-Linear Analysis of a Model for Yeast Cell Communication.
\newblock {\em ESAIM: Mathematical Modelling and Numerical Analysis} {\bf
  2020}, {\em 54},~619--648.
\newblock
  doi:{\changeurlcolor{black}\href{https://doi.org/10.1051/m2an/2019065}{\detokenize{10.1051/m2an/2019065}}}.

\bibitem[Goehring \em{et~al.}(2011)Goehring, Trong, Bois, Chowdhury, Nicola,
  Hyman, and Grill]{Goehring.etal2011}
Goehring, N.W.; Trong, P.K.; Bois, J.S.; Chowdhury, D.; Nicola, E.M.; Hyman,
  A.A.; Grill, S.W.
\newblock Polarization of {{PAR Proteins}} by {{Advective Triggering}} of a
  {{Pattern}}-{{Forming System}}.
\newblock {\em Science} {\bf 2011}, {\em 334},~1137--1141.
\newblock
  doi:{\changeurlcolor{black}\href{https://doi.org/10.1126/science.1208619}{\detokenize{10.1126/science.1208619}}}.

\bibitem[Illukkumbura \em{et~al.}(2020)Illukkumbura, Bland, and
  Goehring]{Illukkumbura.etal2020}
Illukkumbura, R.; Bland, T.; Goehring, N.W.
\newblock Patterning and Polarization of Cells by Intracellular Flows.
\newblock {\em Current Opinion in Cell Biology} {\bf 2020}, {\em 62},~123--134.
\newblock
  doi:{\changeurlcolor{black}\href{https://doi.org/10.1016/j.ceb.2019.10.005}{\detokenize{10.1016/j.ceb.2019.10.005}}}.

\bibitem[Gross \em{et~al.}(2019)Gross, Kumar, Goehring, Bois, Hoege,
  J{\"u}licher, and Grill]{Gross.etal2019}
Gross, P.; Kumar, K.V.; Goehring, N.W.; Bois, J.S.; Hoege, C.; J{\"u}licher,
  F.; Grill, S.W.
\newblock Guiding Self-Organized Pattern Formation in Cell Polarity
  Establishment.
\newblock {\em Nature Physics} {\bf 2019}, {\em 15},~293--300.
\newblock
  doi:{\changeurlcolor{black}\href{https://doi.org/10.1038/s41567-018-0358-7}{\detokenize{10.1038/s41567-018-0358-7}}}.

\bibitem[Ivanov and Mizuuchi(2010)]{Ivanov.Mizuuchi2010}
Ivanov, V.; Mizuuchi, K.
\newblock Multiple Modes of Interconverting Dynamic Pattern Formation by
  Bacterial Cell Division Proteins.
\newblock {\em Proceedings of the National Academy of Sciences} {\bf 2010},
  {\em 107},~8071--8078.
\newblock
  doi:{\changeurlcolor{black}\href{https://doi.org/10.1073/pnas.0911036107}{\detokenize{10.1073/pnas.0911036107}}}.

\bibitem[Vecchiarelli \em{et~al.}(2014)Vecchiarelli, Li, Mizuuchi, and
  Mizuuchi]{Vecchiarelli.etal2014}
Vecchiarelli, A.G.; Li, M.; Mizuuchi, M.; Mizuuchi, K.
\newblock Differential Affinities of {{MinD}} and {{MinE}} to Anionic
  Phospholipid Influence {{Min}} Patterning Dynamics {\emph{in Vitro}}:
  {{Flow}} and Lipid Composition Effects on {{Min}} Patterning.
\newblock {\em Molecular Microbiology} {\bf 2014}, {\em 93},~453--463.
\newblock
  doi:{\changeurlcolor{black}\href{https://doi.org/10.1111/mmi.12669}{\detokenize{10.1111/mmi.12669}}}.

\bibitem[Grill \em{et~al.}(2001)Grill, G{\"o}nczy, Stelzer, and
  Hyman]{Grill.etal2001}
Grill, S.W.; G{\"o}nczy, P.; Stelzer, E.H.K.; Hyman, A.A.
\newblock Polarity Controls Forces Governing Asymmetric Spindle Positioning in
  the {{Caenorhabditis}} Elegans Embryo.
\newblock {\em Nature} {\bf 2001}, {\em 409},~630--633.
\newblock
  doi:{\changeurlcolor{black}\href{https://doi.org/10.1038/35054572}{\detokenize{10.1038/35054572}}}.

\bibitem[Munro \em{et~al.}(2004)Munro, Nance, and Priess]{Munro.etal2004}
Munro, E.; Nance, J.; Priess, J.R.
\newblock Cortical {{Flows Powered}} by {{Asymmetrical Contraction Transport
  PAR Proteins}} to {{Establish}} and {{Maintain Anterior}}-{{Posterior
  Polarity}} in the {{Early C}}. Elegans {{Embryo}}.
\newblock {\em Developmental Cell} {\bf 2004}, {\em 7},~413--424.
\newblock
  doi:{\changeurlcolor{black}\href{https://doi.org/10.1016/j.devcel.2004.08.001}{\detokenize{10.1016/j.devcel.2004.08.001}}}.

\bibitem[Hecht \em{et~al.}(2009)Hecht, Rappel, and Levine]{Hecht.etal2009}
Hecht, I.; Rappel, W.J.; Levine, H.
\newblock Determining the Scale of the {{Bicoid}} Morphogen Gradient.
\newblock {\em Proceedings of the National Academy of Sciences} {\bf 2009},
  {\em 106},~1710--1715.
\newblock
  doi:{\changeurlcolor{black}\href{https://doi.org/10.1073/pnas.0807655106}{\detokenize{10.1073/pnas.0807655106}}}.

\bibitem[Mayer \em{et~al.}(2010)Mayer, Depken, Bois, J{\"u}licher, and
  Grill]{Mayer.etal2010}
Mayer, M.; Depken, M.; Bois, J.S.; J{\"u}licher, F.; Grill, S.W.
\newblock Anisotropies in Cortical Tension Reveal the Physical Basis of
  Polarizing Cortical Flows.
\newblock {\em Nature} {\bf 2010}, {\em 467},~617--621.
\newblock
  doi:{\changeurlcolor{black}\href{https://doi.org/10.1038/nature09376}{\detokenize{10.1038/nature09376}}}.

\bibitem[Goldstein and {van de Meent}(2015)]{Goldstein.vandeMeent2015}
Goldstein, R.E.; {van de Meent}, J.W.
\newblock A Physical Perspective on Cytoplasmic Streaming.
\newblock {\em Interface Focus} {\bf 2015}, {\em 5},~20150030.
\newblock
  doi:{\changeurlcolor{black}\href{https://doi.org/10.1098/rsfs.2015.0030}{\detokenize{10.1098/rsfs.2015.0030}}}.

\bibitem[Mogilner and Manhart(2018)]{Mogilner.Manhart2018}
Mogilner, A.; Manhart, A.
\newblock Intracellular {{Fluid Mechanics}}: {{Coupling Cytoplasmic Flow}} with
  {{Active Cytoskeletal Gel}}.
\newblock {\em Annual Review of Fluid Mechanics} {\bf 2018}, {\em
  50},~347--370.
\newblock
  doi:{\changeurlcolor{black}\href{https://doi.org/10.1146/annurev-fluid-010816-060238}{\detokenize{10.1146/annurev-fluid-010816-060238}}}.

\bibitem[Bischof \em{et~al.}(2017)Bischof, Brand, Somogyi, M{\'a}jer, Thome,
  Mori, Schwarz, and L{\'e}n{\'a}rt]{Bischof.etal2017}
Bischof, J.; Brand, C.A.; Somogyi, K.; M{\'a}jer, I.; Thome, S.; Mori, M.;
  Schwarz, U.S.; L{\'e}n{\'a}rt, P.
\newblock A Cdk1 Gradient Guides Surface Contraction Waves in Oocytes.
\newblock {\em Nature Communications} {\bf 2017}, {\em 8},~1--10.
\newblock
  doi:{\changeurlcolor{black}\href{https://doi.org/10.1038/s41467-017-00979-6}{\detokenize{10.1038/s41467-017-00979-6}}}.

\bibitem[Koslover \em{et~al.}(2017)Koslover, Chan, and
  Theriot]{Koslover.etal2017}
Koslover, E.F.; Chan, C.K.; Theriot, J.A.
\newblock Cytoplasmic {{Flow}} and {{Mixing Due}} to {{Deformation}} of
  {{Motile Cells}}.
\newblock {\em Biophysical Journal} {\bf 2017}, {\em 113},~2077--2087.
\newblock
  doi:{\changeurlcolor{black}\href{https://doi.org/10.1016/j.bpj.2017.09.009}{\detokenize{10.1016/j.bpj.2017.09.009}}}.

\bibitem[Klughammer \em{et~al.}(2018)Klughammer, Bischof, Schnellb{\"a}cher,
  Callegari, L{\'e}n{\'a}rt, and Schwarz]{Klughammer.etal2018}
Klughammer, N.; Bischof, J.; Schnellb{\"a}cher, N.D.; Callegari, A.;
  L{\'e}n{\'a}rt, P.; Schwarz, U.
\newblock Cytoplasmic Flows in Starfish Oocytes Are Fully Determined by
  Cortical Contractions.
\newblock {\em PLoS computational biology} {\bf 2018}, {\em 14},~e1006588.
\newblock
  doi:{\changeurlcolor{black}\href{https://doi.org/https://doi.org/10.1371/journal.pcbi.1006588}{\detokenize{https://doi.org/10.1371/journal.pcbi.1006588}}}.

\bibitem[Mori \em{et~al.}(2008)Mori, Jilkine, and
  {Edelstein-Keshet}]{Mori.etal2008}
Mori, Y.; Jilkine, A.; {Edelstein-Keshet}, L.
\newblock Wave-{{Pinning}} and {{Cell Polarity}} from a {{Bistable
  Reaction}}-{{Diffusion System}}.
\newblock {\em Biophysical Journal} {\bf 2008}, {\em 94},~3684--3697.
\newblock
  doi:{\changeurlcolor{black}\href{https://doi.org/10.1529/biophysj.107.120824}{\detokenize{10.1529/biophysj.107.120824}}}.

\bibitem[Jilkine and {Edelstein-Keshet}(2011)]{Jilkine.Edelstein-Keshet2011}
Jilkine, A.; {Edelstein-Keshet}, L.
\newblock A {{Comparison}} of {{Mathematical Models}} for {{Polarization}} of
  {{Single Eukaryotic Cells}} in {{Response}} to {{Guided Cues}}.
\newblock {\em PLoS Computational Biology} {\bf 2011}, {\em 7},~e1001121.
\newblock
  doi:{\changeurlcolor{black}\href{https://doi.org/10.1371/journal.pcbi.1001121}{\detokenize{10.1371/journal.pcbi.1001121}}}.

\bibitem[Diegmiller \em{et~al.}(2018)Diegmiller, Montanelli, Muratov, and
  Shvartsman]{Diegmiller.etal2018}
Diegmiller, R.; Montanelli, H.; Muratov, C.B.; Shvartsman, S.Y.
\newblock Spherical {{Caps}} in {{Cell Polarization}}.
\newblock {\em Biophysical Journal} {\bf 2018}, {\em 115},~26--30.
\newblock
  doi:{\changeurlcolor{black}\href{https://doi.org/10.1016/j.bpj.2018.05.033}{\detokenize{10.1016/j.bpj.2018.05.033}}}.

\bibitem[Chiou \em{et~al.}(2018)Chiou, Ramirez, Elston, Witelski, Schaeffer,
  and Lew]{Chiou.etal2018}
Chiou, J.G.; Ramirez, S.A.; Elston, T.C.; Witelski, T.P.; Schaeffer, D.G.; Lew,
  D.J.
\newblock Principles That Govern Competition or Co-Existence in
  {{Rho}}-{{GTPase}} Driven Polarization.
\newblock {\em PLOS Computational Biology} {\bf 2018}, {\em 14},~e1006095.
\newblock
  doi:{\changeurlcolor{black}\href{https://doi.org/10.1371/journal.pcbi.1006095}{\detokenize{10.1371/journal.pcbi.1006095}}}.

\bibitem[Brauns \em{et~al.}(2018)Brauns, Halatek, and Frey]{Brauns.etal2018}
Brauns, F.; Halatek, J.; Frey, E.
\newblock Phase-Space Geometry of Reaction--Diffusion Dynamics.
\newblock {\em arXiv:1812.08684} {\bf 2018},
  \href{http://xxx.lanl.gov/abs/1812.08684}{{\normalfont [1812.08684]}}.

\bibitem[Otsuji \em{et~al.}(2007)Otsuji, Ishihara, Co, Kaibuchi, Mochizuki, and
  Kuroda]{Otsuji.etal2007}
Otsuji, M.; Ishihara, S.; Co, C.; Kaibuchi, K.; Mochizuki, A.; Kuroda, S.
\newblock A {{Mass Conserved Reaction}}\textendash{{Diffusion System Captures
  Properties}} of {{Cell Polarity}}.
\newblock {\em PLoS Computational Biology} {\bf 2007}, {\em 3},~e108.
\newblock
  doi:{\changeurlcolor{black}\href{https://doi.org/10.1371/journal.pcbi.0030108}{\detokenize{10.1371/journal.pcbi.0030108}}}.

\bibitem[Goryachev and Pokhilko(2008)]{Goryachev.Pokhilko2008}
Goryachev, A.B.; Pokhilko, A.V.
\newblock Dynamics of {{Cdc42}} Network Embodies a {{Turing}}-Type Mechanism of
  Yeast Cell Polarity.
\newblock {\em FEBS Letters} {\bf 2008}, {\em 582},~1437--1443.
\newblock
  doi:{\changeurlcolor{black}\href{https://doi.org/10.1016/j.febslet.2008.03.029}{\detokenize{10.1016/j.febslet.2008.03.029}}}.

\bibitem[Trong \em{et~al.}(2014)Trong, Nicola, Goehring, Kumar, and
  Grill]{Trong.etal2014}
Trong, P.K.; Nicola, E.M.; Goehring, N.W.; Kumar, K.V.; Grill, S.W.
\newblock Parameter-Space Topology of Models for Cell Polarity.
\newblock {\em New Journal of Physics} {\bf 2014}, {\em 16},~065009.
\newblock
  doi:{\changeurlcolor{black}\href{https://doi.org/10.1088/1367-2630/16/6/065009}{\detokenize{10.1088/1367-2630/16/6/065009}}}.

\bibitem[Ishihara \em{et~al.}(2007)Ishihara, Otsuji, and
  Mochizuki]{Ishihara.etal2007}
Ishihara, S.; Otsuji, M.; Mochizuki, A.
\newblock Transient and Steady State of Mass-Conserved Reaction-Diffusion
  Systems.
\newblock {\em Physical Review E} {\bf 2007}, {\em 75},~015203.
\newblock
  doi:{\changeurlcolor{black}\href{https://doi.org/10.1103/PhysRevE.75.015203}{\detokenize{10.1103/PhysRevE.75.015203}}}.

\bibitem[Wigbers \em{et~al.}(2020)Wigbers, Brauns, Hermann, and
  Frey]{Wigbers.etal2020}
Wigbers, M.C.; Brauns, F.; Hermann, T.; Frey, E.
\newblock Pattern Localization to a Domain Edge.
\newblock {\em Physical Review E} {\bf 2020}, {\em 101},~022414.
\newblock
  doi:{\changeurlcolor{black}\href{https://doi.org/10.1103/PhysRevE.101.022414}{\detokenize{10.1103/PhysRevE.101.022414}}}.

\bibitem[Halatek \em{et~al.}(2018)Halatek, Brauns, and Frey]{Halatek.etal2018}
Halatek, J.; Brauns, F.; Frey, E.
\newblock Self-Organization Principles of Intracellular Pattern Formation.
\newblock {\em Philosophical Transactions of the Royal Society B: Biological
  Sciences} {\bf 2018}, {\em 373},~20170107.
\newblock
  doi:{\changeurlcolor{black}\href{https://doi.org/10.1098/rstb.2017.0107}{\detokenize{10.1098/rstb.2017.0107}}}.

\bibitem[Allen and Allen(1978)]{Allen.Allen1978}
Allen, N.S.; Allen, R.D.
\newblock Cytoplasmic {{Streaming}} in {{Green Plants}}.
\newblock {\em Annual Review of Biophysics and Bioengineering} {\bf 1978}, {\em
  7},~497--526.
\newblock
  doi:{\changeurlcolor{black}\href{https://doi.org/10.1146/annurev.bb.07.060178.002433}{\detokenize{10.1146/annurev.bb.07.060178.002433}}}.

\bibitem[Siero \em{et~al.}(2015)Siero, Doelman, Eppinga, Rademacher, Rietkerk,
  and Siteur]{Siero.etal2015}
Siero, E.; Doelman, A.; Eppinga, M.; Rademacher, J.D.; Rietkerk, M.; Siteur, K.
\newblock Striped Pattern Selection by Advective Reaction-Diffusion Systems:
  {{Resilience}} of Banded Vegetation on Slopes.
\newblock {\em Chaos: An Interdisciplinary Journal of Nonlinear Science} {\bf
  2015}, {\em 25},~036411.

\bibitem[Perumpanani \em{et~al.}(1995)Perumpanani, Sherratt, and
  Maini]{Perumpanani.etal1995}
Perumpanani, A.J.; Sherratt, J.A.; Maini, P.K.
\newblock Phase Differences in Reaction-Diffusion-Advection Systems and
  Applications to Morphogenesis.
\newblock {\em IMA Journal of Applied Mathematics} {\bf 1995}, {\em
  55},~19--33.
\newblock
  doi:{\changeurlcolor{black}\href{https://doi.org/10.1093/imamat/55.1.19}{\detokenize{10.1093/imamat/55.1.19}}}.

\bibitem[Samuelson \em{et~al.}(2019)Samuelson, Singer, Weinburd, and
  Scheel]{Samuelson.etal2019}
Samuelson, R.; Singer, Z.; Weinburd, J.; Scheel, A.
\newblock {Advection and Autocatalysis as Organizing Principles for Banded
  Vegetation Patterns}.
\newblock {\em Journal of Nonlinear Science} {\bf 2019}, {\em 29},~255--285.
\newblock
  doi:{\changeurlcolor{black}\href{https://doi.org/10.1007/s00332-018-9486-6}{\detokenize{10.1007/s00332-018-9486-6}}}.

\bibitem[Mat(2019)]{Mathematica}
Mathematica.
\newblock Wolfram Research, Inc.,  2019.

\bibitem[Brauns \em{et~al.}(2020)Brauns, Weyer, Halatek, Yoon, and
  Frey]{Brauns.etal2020}
Brauns, F.; Weyer, H.; Halatek, J.; Yoon, J.; Frey, E.
\newblock Coarsening in (Nearly) Mass-Conserving Two-Component Reaction
  Diffusion Systems.
\newblock {\em arXiv:2005.01495 [nlin, physics:physics]} {\bf 2020},
  \href{http://xxx.lanl.gov/abs/2005.01495}{{\normalfont [arXiv:nlin,
  physics:physics/2005.01495]}}.

\bibitem[Fivaz \em{et~al.}(2008)Fivaz, Bandara, Inoue, and
  Meyer]{Fivaz.etal2008}
Fivaz, M.; Bandara, S.; Inoue, T.; Meyer, T.
\newblock Robust {{Neuronal Symmetry Breaking}} by {{Ras}}-{{Triggered Local
  Positive Feedback}}.
\newblock {\em Current Biology} {\bf 2008}, {\em 18},~44--50.
\newblock
  doi:{\changeurlcolor{black}\href{https://doi.org/10.1016/j.cub.2007.11.051}{\detokenize{10.1016/j.cub.2007.11.051}}}.

\bibitem[Wang \em{et~al.}(2013)Wang, Ku, Zhang, Artyukhin, Weiner, Wu, and
  Altschuler]{Wang.etal2013}
Wang, Y.; Ku, C.J.; Zhang, E.R.; Artyukhin, A.B.; Weiner, O.D.; Wu, L.F.;
  Altschuler, S.J.
\newblock Identifying {{Network Motifs}} That {{Buffer Front}}-to-{{Back
  Signaling}} in {{Polarized Neutrophils}}.
\newblock {\em Cell Reports} {\bf 2013}, {\em 3},~1607--1616.
\newblock
  doi:{\changeurlcolor{black}\href{https://doi.org/10.1016/j.celrep.2013.04.009}{\detokenize{10.1016/j.celrep.2013.04.009}}}.

\bibitem[Halatek and Frey(2018)]{Halatek.Frey2018}
Halatek, J.; Frey, E.
\newblock Rethinking Pattern Formation in Reaction\textendash{}Diffusion
  Systems.
\newblock {\em Nature Physics} {\bf 2018}, {\em 14},~507--514.
\newblock
  doi:{\changeurlcolor{black}\href{https://doi.org/10.1038/s41567-017-0040-5}{\detokenize{10.1038/s41567-017-0040-5}}}.

\bibitem[Wang \em{et~al.}(2017)Wang, Low, Nishimura, Gole, Yu, and
  Motegi]{Wang.etal2017}
Wang, S.C.; Low, T.Y.F.; Nishimura, Y.; Gole, L.; Yu, W.; Motegi, F.
\newblock Cortical Forces and {{CDC}}-42 Control Clustering of {{PAR}} Proteins
  for {{Caenorhabditis}} Elegans Embryonic Polarization.
\newblock {\em Nature Cell Biology} {\bf 2017}, {\em 19},~988--995.
\newblock
  doi:{\changeurlcolor{black}\href{https://doi.org/10.1038/ncb3577}{\detokenize{10.1038/ncb3577}}}.

\bibitem[Mittasch \em{et~al.}(2018)Mittasch, Gross, Nestler, Fritsch, Iserman,
  Kar, Munder, Voigt, Alberti, Grill, and Kreysing]{Mittasch.etal2018}
Mittasch, M.; Gross, P.; Nestler, M.; Fritsch, A.W.; Iserman, C.; Kar, M.;
  Munder, M.; Voigt, A.; Alberti, S.; Grill, S.W.; Kreysing, M.
\newblock Non-Invasive Perturbations of Intracellular Flow Reveal Physical
  Principles of Cell Organization.
\newblock {\em Nature Cell Biology} {\bf 2018}, {\em 20},~344--351.
\newblock
  doi:{\changeurlcolor{black}\href{https://doi.org/10.1038/s41556-017-0032-9}{\detokenize{10.1038/s41556-017-0032-9}}}.

\bibitem[Loose \em{et~al.}(2008)Loose, {Fischer-Friedrich}, Ries, Kruse, and
  Schwille]{Loose.etal2008}
Loose, M.; {Fischer-Friedrich}, E.; Ries, J.; Kruse, K.; Schwille, P.
\newblock Spatial {{Regulators}} for {{Bacterial Cell Division
  Self}}-{{Organize}} into {{Surface Waves}} in {{Vitro}}.
\newblock {\em Science} {\bf 2008}, {\em 320},~789--792.
\newblock
  doi:{\changeurlcolor{black}\href{https://doi.org/10.1126/science.1154413}{\detokenize{10.1126/science.1154413}}}.

\bibitem[Glock \em{et~al.}(2019)Glock, Ramm, Heermann, Kretschmer, Schweizer,
  M{\"u}cksch, Alag{\"o}z, and Schwille]{Glock.etal2019}
Glock, P.; Ramm, B.; Heermann, T.; Kretschmer, S.; Schweizer, J.; M{\"u}cksch,
  J.; Alag{\"o}z, G.; Schwille, P.
\newblock Stationary {{Patterns}} in a {{Two}}-{{Protein Reaction}}-{{Diffusion
  System}}.
\newblock {\em ACS Synthetic Biology} {\bf 2019}, {\em 8},~148--157.
\newblock
  doi:{\changeurlcolor{black}\href{https://doi.org/10.1021/acssynbio.8b00415}{\detokenize{10.1021/acssynbio.8b00415}}}.

\bibitem[Huang \em{et~al.}(2003)Huang, Meir, and Wingreen]{Huang.etal2003}
Huang, K.C.; Meir, Y.; Wingreen, N.S.
\newblock Dynamic Structures in {{Escherichia}} Coli: {{Spontaneous}} Formation
  of {{MinE}} Rings and {{MinD}} Polar Zones.
\newblock {\em Proceedings of the National Academy of Sciences} {\bf 2003},
  {\em 100},~12724--12728.
\newblock
  doi:{\changeurlcolor{black}\href{https://doi.org/10.1073/pnas.2135445100}{\detokenize{10.1073/pnas.2135445100}}}.

\bibitem[Halatek and Frey(2012)]{Halatek.Frey2012}
Halatek, J.; Frey, E.
\newblock Highly {{Canalized MinD Transfer}} and {{MinE Sequestration Explain}}
  the {{Origin}} of {{Robust MinCDE}}-{{Protein Dynamics}}.
\newblock {\em Cell Reports} {\bf 2012}, {\em 1},~741--752.
\newblock
  doi:{\changeurlcolor{black}\href{https://doi.org/10.1016/j.celrep.2012.04.005}{\detokenize{10.1016/j.celrep.2012.04.005}}}.

\bibitem[Deneke \em{et~al.}(2019)Deneke, Puliafito, Krueger, Narla, De~Simone,
  Primo, Vergassola, De~Renzis, and Di~Talia]{Deneke.etal2019}
Deneke, V.E.; Puliafito, A.; Krueger, D.; Narla, A.V.; De~Simone, A.; Primo,
  L.; Vergassola, M.; De~Renzis, S.; Di~Talia, S.
\newblock Self-{{Organized Nuclear Positioning Synchronizes}} the {{Cell
  Cycle}} in {{Drosophila Embryos}}.
\newblock {\em Cell} {\bf 2019}, {\em 177},~925--941.e17.
\newblock
  doi:{\changeurlcolor{black}\href{https://doi.org/10.1016/j.cell.2019.03.007}{\detokenize{10.1016/j.cell.2019.03.007}}}.

\bibitem[Iden and Collard(2008)]{Iden.Collard2008}
Iden, S.; Collard, J.G.
\newblock Crosstalk between Small {{GTPases}} and Polarity Proteins in Cell
  Polarization.
\newblock {\em Nature Reviews Molecular Cell Biology} {\bf 2008}, {\em
  9},~846--859.
\newblock
  doi:{\changeurlcolor{black}\href{https://doi.org/10.1038/nrm2521}{\detokenize{10.1038/nrm2521}}}.

\bibitem[Bois \em{et~al.}(2011)Bois, J{\"u}licher, and Grill]{Bois.etal2011}
Bois, J.S.; J{\"u}licher, F.; Grill, S.W.
\newblock Pattern {{Formation}} in {{Active Fluids}}.
\newblock {\em Physical Review Letters} {\bf 2011}, {\em 106},~028103.
\newblock
  doi:{\changeurlcolor{black}\href{https://doi.org/10.1103/PhysRevLett.106.028103}{\detokenize{10.1103/PhysRevLett.106.028103}}}.

\bibitem[Radszuweit \em{et~al.}(2013)Radszuweit, Alonso, Engel, and
  B{\"a}r]{Radszuweit.etal2013}
Radszuweit, M.; Alonso, S.; Engel, H.; B{\"a}r, M.
\newblock Intracellular {{Mechanochemical Waves}} in an {{Active Poroelastic
  Model}}.
\newblock {\em Physical Review Letters} {\bf 2013}, {\em 110},~138102.
\newblock
  doi:{\changeurlcolor{black}\href{https://doi.org/10.1103/PhysRevLett.110.138102}{\detokenize{10.1103/PhysRevLett.110.138102}}}.

\bibitem[Kumar \em{et~al.}(2014)Kumar, Bois, J{\"u}licher, and
  Grill]{Kumar.etal2014}
Kumar, K.V.; Bois, J.S.; J{\"u}licher, F.; Grill, S.W.
\newblock Pulsatory {{Patterns}} in {{Active Fluids}}.
\newblock {\em Physical Review Letters} {\bf 2014}, {\em 112}.
\newblock
  doi:{\changeurlcolor{black}\href{https://doi.org/10.1103/PhysRevLett.112.208101}{\detokenize{10.1103/PhysRevLett.112.208101}}}.

\bibitem[Nakagaki \em{et~al.}(1999)Nakagaki, Yamada, and
  Ito]{Nakagaki.etal1999}
Nakagaki, T.; Yamada, H.; Ito, M.
\newblock Reaction\textendash{{Diffusion}}\textendash{{Advection Model}} for
  {{Pattern Formation}} of {{Rhythmic Contraction}} in a {{Giant Amoeboid
  Cell}} of {{thePhysarumPlasmodium}}.
\newblock {\em Journal of Theoretical Biology} {\bf 1999}, {\em 197},~497--506.
\newblock
  doi:{\changeurlcolor{black}\href{https://doi.org/10.1006/jtbi.1998.0890}{\detokenize{10.1006/jtbi.1998.0890}}}.

\end{thebibliography}


\end{document}